\documentclass[usenatbib]{mn2e}
\usepackage{natbibmnfix,graphicx,times, amsmath}

\def\myputfigure#1#2#3#4#5%
{\vskip#5pt\makebox[0pt]{\hskip#2in
\includegraphics[width=#3\textwidth]{#1}}\vskip#4pt\hfill}

\newcommand\lsim{\mathrel{\rlap{\lower4pt\hbox{\hskip1pt$\sim$}}
        \raise1pt\hbox{$<$}}}
\newcommand\gsim{\mathrel{\rlap{\lower4pt\hbox{\hskip1pt$\sim$}}
        \raise1pt\hbox{$>$}}}

\newcommand{\avenf}{\bar{x}_{\rm HI}}

\newcommand{\hs}{\hspace{1mm}}

\newcommand{\Msun}{M_\odot}
\newcommand{\Tvir}{T_{\rm vir}}
\newcommand{\fcoll}{f_{\rm gas}}

\newcommand{\Mmin}{M_{\rm min}}

\newcommand{\zon}{z_{\rm on}}
\newcommand{\lmfp}{l_{\rm mfp}}
\newcommand{\toteff}{\epsilon_{\rm fid}}

\newcommand{\flux}{f({\bf x}, z)}
\newcommand{\Jcrit}{J^{\rm crit}_{21}(M, z)}
\newcommand{\kmps}{\rm km~s^{-1}}

\newcommand{\apjl}{ApJ}
\newcommand{\apjs}{ApJS}
\newcommand{\aap}{A\&A}

\begin{document}

%\submitted{Submitted to the MNRAS}

\title[UV Radiative Feedback During the Advanced Stages of Reionization]{UV Radiative Feedback During the Advanced Stages of Reionization}

\author[Mesinger \& Dijkstra]{Andrei Mesinger$^1$\thanks{Email: mesinger@astro.ucla.edu} \& Mark Dijkstra$^2$ \\
$^1$Department of Physics and Astronomy, UCLA, Los Angeles, CA 90095-1562, USA\\
$^2$Harvard-Smithsonian Center for Astrophysics, 60 Garden Street, Cambridge, MA 02138, USA}

\voffset-.6in

\maketitle

\begin{abstract}
The ionizing ultraviolet background (UVB) during reionization can suppress the gas content of low-mass galaxies, even those capable of efficient atomic cooling (i.e. with virial temperatures $\Tvir \gsim 10^4$ K). This negative radiative feedback mechanism can thus reduce the star formation efficiencies of these halos, which may delay the completion of reionization. In this work, we explore the importance of UV radiative feedback on $\Tvir \gsim 10^4$ K halos during the middle and late stages of reionization.  We do not try to self-consistently model reionization; instead, we explore a large parameter space in an attempt to draw general, robust conclusions. 
We use a tiered approach. Using 1-D hydrodynamical simulations, we model the ability of gas to collapse onto halos of various masses under UVBs of various intensities.  We then generate realistic, parametrized maps of the inhomogeneous UVB, using large-scale semi-numeric simulations. By combining these results, we find that under all reasonably conservative scenarios, UV feedback on atomically-cooled halos is not strong enough to notably delay the bulk of reionization. Such a delay is only likely if ionizing efficiencies of $z\gsim10$ sources are much higher ($\sim$ two orders of magnitude) than $z\sim6$ data seem to imply.  Towards the end of reionization, star formation can be quenched only in halos in a narrow mass range close to the atomic-cooling threshold. This result depends only weakly on the intensity of the UVB: quenching star formation in halos just twice as massive requires an order of magnitude increase in source ionizing efficiencies. This implies that the natural time-scale for the bulk of reionization is the growth of the global collapsed fraction contained in $\Tvir\gsim 10^4$ K halos.
Thus the likely reionization scenario would involve a small HII filling factor ``tail'' extending to high redshifts, governed by more complicated feedback on $\Tvir \lsim 10^4$ K objects, followed by a period of relatively rapid evolution in the HII filling factor. Furthermore, our results underscore the importance of extended dynamical ranges when modeling reionization.  Simulations must be capable of resolving halos with mass $\gsim 10^8 \Msun$, even when modeling the late stages of reionization, while at the same time being large enough to capture HII regions several tens of Mpc in size.
\end{abstract}


\section{Introduction}
\label{sec:intro}

The epoch of reionization, when light from early generations of astrophysical objects began flowing through the intergalactic medium (IGM), remains one of the most compelling frontiers of modern cosmology, offering a wealth of information about cosmological structure formation and physical processes in the early universe.  Only recently have we begun to gather clues concerning this epoch.  The presence of flux in the Lyman line absorption regions of $z\lsim6$ quasars discovered in the Sloan Digital Sky Survey (SDSS) indicates that reionization is mostly completed by this redshift (though note that large-scale fluctuations in the ionization field can yield long ionized sightlines even before the completion of reionization, e.g. \citealt{Lidz07}).  Spectra from some higher redshift quasars have even been interpreted as evidence of a partially neutral IGM \citep{MH04, WL04_nf, Fan06, MH07}, though most of these interpretations are controversial \citep{LOF06, BH07, Maselli07, BRS07, MF08damp}. The {\it Wilkinson Microwave Anisotropy Probe} ({\it WMAP}) measured a Thomson scattering optical depth of $\tau_e = 0.087\pm0.017$ \citep{Dunkley08}.  Assuming instantaneous reionization, this value of $\tau_e$ corresponds to a reionization redshift of $z = 11.0 \pm 1.4$.  However, this is only an integrated measurement and can tell us very little about the evolution of reionization.

With the interpretation of even the sparse existing data being subtle and complicated, much effort has been invested in improving the modeling of reionization.  We think we can accurately model the abundances and clustering properties of dark matter halos at high redshift.  Combined with some analytic relation of halo mass $\rightarrow$ effective ionizing photon emissivity, one can get an estimate of how the global neutral fraction evolves; this technique has been used is countless reionization studies.  Aside from the seemingly magical ``$\rightarrow$'' in the previous sentence, one of the main additional uncertainties are the feedback processes: how do sources impact the current and future generations of sources?  Answering this question is non-trivial, especially in the early universe.  Radiative, chemical and mechanical feedback can effect the ionizing efficiencies of the first, highly biased sources.  Molecular hydrogen (H$_2$) cooling can allow very small (with virial temperatures, $\Tvir$, of several hundred Kelvin) halos to host astrophysical sources.  At later times (though still prior to the bulk of reionization, \citealt{HRL97}), as the H$_2$ dissociative background builds-up and the contribution of molecularly-cooled halos declines, feedback enters the regime of atomically-cooled halos, $\Tvir \gsim 10^4$ K.  This regime is more straightforward to model as negative radiative feedback emerges as a single dominant mechanism, especially on large scales.

In the presence of an ionizing background radiation, the IGM is
photo--heated to a temperature of $\gsim 10^4$K, raising the
cosmological Jeans mass, which could suppress gas accretion onto
small-mass halos (e.g. \citealt{Efstathiou92, SGB94, TW96, HG97}).
 Early work on this subject (so-called ``Jeans mass filtering'') suggested that an ionizing background would completely
suppress star formation in low--redshift ``dwarf galaxy'' halos with
circular velocities $v_{\rm circ} \lsim~35~\kmps$, and partially
suppress star--formation in halos with 35 $\kmps$ $\lsim$ $v_{\rm
circ}$ $\lsim$ 100 $\kmps$ \citep{TW96}.  Many reionization studies assume prescriptions of gas suppression based on these results (e.g. \citealt{RS03, OM04, WC07, Iliev07, WBH08}).  However, more recent studies \citep{KI00, Dijkstra04} find that at $z \gsim 3$, more compact halo profiles, increased cooling efficiencies, and shorter exposure times to the ultraviolet background (UVB) could lessen the importance of negative radiative feedback.  We expand on these works by exploring a wider parameter space, placed in a broader context of an inhomogenious reionization with a patchy UVB due to both source clustering and modulation by HII regions.  \citet{Gnedin00filter} studied the effects of such radiative feedback in cosmological simulations; however, these simulation boxes were of necessity very small and only included a single reionization model.  

{\it The purpose of this paper is to explore the importance of UV radiative feedback during the middle and late stages of reionization}. We do not attempt to self-consistently model feedback during reionization; such a thing is beyond the capacity of current simulations, especially given our poor understanding of the first generations of astrophysical sources and their environments.  Instead we statistically present the effects of an inhomogeneous UVB on the suppression of gas content in low-mass galaxies {\it capable of atomic cooling}, i.e. $\Tvir \gsim 10^4$ K.  We do this using a tiered approach: using numerical simulations \citep{TW95, Dijkstra04} to calibrate very large scale, high resolution ``semi-numerical'' simulations \citep{MF07}.  In order to keep things general, we explore wide swaths of parameter space keeping assumptions minimal.

This paper is organized as follows. In \S \ref{sec:mark_sims} and \S \ref{sec:my_sims}, we describe our hydrodynamic collapse simulations and semi-numerical cosmological simulations, respectively.  In \S \ref{sec:fcoll} we present the results from our collapse simulations covering a wide range of parameter choices, while in \S \ref{sec:dist} we present parametrized distributions of UV fluxes from our cosmological simulations. In \S \ref{sec:feedback} we combine these results to quantify the importance of UV feedback during reionization.  In \S \ref{sec:ass} we discuss the assumptions and uncertainties in our approach. Finally, in \S \ref{sec:conc} we present our conclusions.

We quote all quantities in comoving units, with the exception of flux, where we denote proper units with a prefix 'p'. We adopt the background cosmological parameters ($\Omega_\Lambda$, $\Omega_{\rm M}$, $\Omega_b$, $n$, $\sigma_8$, $H_0$) = (0.76, 0.24, 0.0407, 0.96, 0.76, 72 km s$^{-1}$ Mpc$^{-1}$), matching the three--year results of the {\it WMAP} satellite \citep{Spergel07}, which in turn are consistent with the recent five-year data release \citep{Komatsu08}.

\section{Simulations}
\label{sec:sims}

As we outlined above, we use spherically-symmetric simulations to model the collapse of gas onto halos with masses $M \gsim 10^8 \Msun$.  With these numerical simulations, we are able to determine the amount of gas which collapses onto a halo of mass $M$ at $z$, under the presence of a UVB with specific intensity $J_{21}$ turned on at redshift $\zon$.  We then generate large-scale halo, ionization and parametrized flux fields, using semi-numerical techniques.  Combining these results, we investigate what fraction of halos experience strong negative radiative feedback.  We describe our methodology and numerical techniques in greater detail below.

\subsection{Hydrodynamical Collapse Simulations}
\label{sec:mark_sims}

We investigate the impact of a photoionizing flux on the ability of gas to cool and collapse onto dark matter halos by performing 1-D hydrodynamical simulations which include both gas and dark matter. We use the code that was originally written by \citet{TW95} and modifiedcharacterized to study the impact of photoionization feedback on the formation of high-redshift low-mass galaxies (see Dijkstra et al  2004 for a more detailed description). 

To summarize our calculations: an externally generated UVB impinges upon a spherically symmetric collapsing dark matter halo of mass $M$ that would have collapsed to $r=0$ at redshift $z$, in the absence of gas pressure. The ionizing radiation field is switched on at $z_{\rm on}$, and is characterized by a power-law spectrum of the form $J(\nu)=J_{\rm 21}(\nu/\nu_H)^{-\alpha}\times 10^{-21}$ erg s$^{-1}$ Hz$^{-1}$ pcm$^{-2}$ sr$^{-1}$, where $\nu_H$ is the ionization threshold of hydrogen.  We are interested in the fraction of the gas, $\fcoll \equiv M_{\rm gas}(p)/M_{\rm gas}(p=0)$, (i.e. the mass of collapsed gas divided by the mass of gas which would have collapsed in the absence of pressure) that is instantly available for star formation at $z$. Therefore, we compute the fraction of gas which has collapsed at $z$ (with an associated Hubble time $t_{\rm coll}$) which differs from the calculations presented by e.g. Thoul \& Weinberg (1996), and Dijkstra et al. (2004) who computed the fraction of gas which was able to collapse up to $2t_{\rm coll}$. 

For all runs, we switch on the UVB at $\zon=14$. The 5-yr {\it WMAP} polarization data place a constraint on the reionization redshift (assuming instantaneous reionization) of $z=11.0 \pm 1.4$ \citep{Dunkley08}.  Thus it is highly unlikely that the majority of the universe was ionized before $z\sim14$ (for some interpretations of the less-constrained 3-yr data, see, for example \citealt{HB06, CF06}). Therefore, our choice of $\zon=14$ is conservative, since most halos should be exposed to ionizing radiation at later times and so we overestimate the impact of radiative feedback.

Furthermore, following Dijkstra et al. (2004) we have assumed $\alpha=1.0$, which is typically associated with accreting black holes (e.g. \citealt{Madau04}) and not with stellar sources which would be better characterized with $\alpha\sim5$ \citep{TW96, BL01}.
 Because the simulations ignore self-shielding, lowering $\alpha$ is equivalent to increasing $J_{\rm 21}$ in terms of the photoionization and photoheating rates. Therefore, in our calculations we overestimate the impact of the photoionization feedback (see \S~\ref{sec:ass} for a more extended discussion).

\subsection{Semi-Numerical Cosmological Simulations}
\label{sec:my_sims}

We use an excursion-set approach combined with first-order Lagrangian perturbation theory to efficiently generate density, velocity, halo, and ionization fields at $z=$ 7, 10 and 13.  This ``semi-numerical'' simulation is presented in \citet{MF07}, to which we refer the reader for details. A similar halo-finding scheme has also been presented by \citet{BM96_algo} and a similar scheme to generate ionization fields has been presented by \citet{Zahn07}.

Our simulation box is 100 Mpc on a side, with the final  ionization fields having grid cell sizes of 0.5 Mpc.  Halos with a total mass $M \ge 2 \times 10^7$ $\Msun$ are filtered out of the linear density field using excursion-set theory; however, for this work, we only concern ourselves with halos capable of efficient atomic hydrogen cooling, $\Tvir \gsim 10^4$ K, or $M \gsim 10^8 \Msun$ for the redshifts in question.  Halo locations are then adjusted using first-order Lagrangian perturbation theory and velocity resolution of $\sim0.14$ Mpc.  The resulting halo field matches both the mass function and statistical clustering properties of halos in N-body simulations \citep{MF07}.

In constructing the ionization field, the IGM is modeled as a two-phase medium, comprised of fully ionized and fully neutral regions (this is a fairly accurate assumption at high redshifts preceding the end of reionization, unless the X-ray background is rather strong).  For each halo and density field at our three redshifts of interest, $z=7,10,13$, we generate several ionization fields corresponding to different values of $\avenf$ by varying a single efficiency parameter, $\zeta$, again using the excursion-set approach (c.f. \citealt{FHZ04}).

This fast, semi-numeric approach is thus ideally suited to reionization studies, because we are able to ``resolve" relatively small halos and simultaneously sample a large, representative volume of ionized bubbles.

\subsubsection{Generating the Flux Field}
\label{sec:flux}

For each halo and ionization map, we create a corresponding UV flux field on a 200$^3$ grid, by summing the contributions of halos whose sightlines to the cell of interest go through ionized IGM.  Specifically, we compute the flux of ionizing photons (in units of ionizing photons s$^{-1}$ pcm$^{-2}$) with

\begin{equation}
\label{eq:sum}
\flux= \frac{(1+z)^2}{4 \pi} \sum_{i} \frac{M_i\times \epsilon_{\rm ion}}{|{\bf x} - {\bf x_i}|^2} ~ e^{-|{\bf x} - {\bf x_i}|/\lmfp} ~, 
\end{equation}

\noindent  where ${\bf x}$ is the location of the cell of interest, $M_i$ is the total halo mass, ${\bf x_i}$ is its location, $\lmfp$ is the assumed mean free path (m.f.p.) of ionizing photons in the {\it ionized IGM}, and the factor of $(1+z)^2$ converts the factor $|{\bf x} - {\bf x_i}|^2$ from comoving into proper units.  We work in the Euclidian limit (i.e. the relevant distances are $|{\bf x} - {\bf x_i}| \le$ m.f.p. $\ll$ hubble length, and cosmological corrections can be ignored; see eq. \ref{eq:mfp}).
We assume $\lmfp=20$ Mpc throughout, with one exception that is discussed explicitly in \S~\ref{sec:dist}.  This value is consistent at $z\sim$ 7 with extrapolations from the observed $z\lsim4$ abundances of Lyman limit systems\footnote{Note that such extrapolation predicts lower values for the mean free path than our choice at $z\gsim9$. Thus we likely conservatively overestimate $\lmfp$ for the majority of the pertinent redshift range (see also \citealt{HAM01}).}
 (LLSs; e.g. \citealt{Storrie-Lombardi94, Miralda-Escude03, CFG08}). Our choice is also consistent with the theoretical assumption that the m.f.p. in our ionized regions is dominated by self-shielded minihalos (halos with virial temperatures $\Tvir < 10^4$ K).  Assuming a roughly random distribution of minihalos with number density $n_{\rm mh}$ and radius $R_{\rm mh}$ , this yields a m.f.p. of (c.f. \citealt{FO05}):

\begin{align}
\label{eq:mfp}
\nonumber \lmfp(z) &\approx \frac{1}{\pi ~ n_{\rm mh} R^2_{\rm mh}} \approx \\
&\approx 17 \left( \frac{M_{\rm mh}}{10^6 \Msun} \right)^\frac{1}{3} \left( \frac{0.05}{f_{\rm mh}} \right) \left( \frac{\Delta_{\rm mh}}{18\pi^2} \right)^\frac{2}{3} \left( \frac{\Omega_{\rm M} h^2}{0.12} \right) {\rm Mpc},
\end{align}

\noindent where $M_{\rm mh}$ is the typical minihalo mass, $\Delta_{\rm mh}$ its mean overdensity, and $f_{\rm mh}$ the fraction of matter locked up in minihalos.  As defined above, $f_{\rm mh}$, can vary from 0.02 at $z=13$ to 0.09 at $z=7$.  However, since the derivation of $\lmfp$ and its evolution (existing minihalos can get photoevaporated with time, e.g. \citealt{SIR04}, partially countering the buildup of new minihalos) is extremely uncertain, we assume a redshift-independent $\lmfp$.  We will elaborate more on how $\lmfp$ affects our conclusions below.

Finally, and most importantly, $\epsilon_{\rm ion}$ in eq. (\ref{eq:sum}) denotes the rate at which ionizing photons are released into the IGM by a dark matter halo per unit mass. This number depends on several factors: ({\it i}) the number of ionizing photons that is emitted per stellar mass (in $\Msun$),
 $\epsilon_\gamma$, ({\it ii}) the fraction of ionizing photons that escapes from the halo into the IGM, $f_{\rm esc}$, and ({\it iii}) the star formation rate (SFR) inside the halo. Therefore, we can write:

\begin{equation}
\label{eq:epsion}
\epsilon_{\rm ion}(z)=\epsilon_\gamma f_{\rm esc}\left[\frac{\Omega_b}{\Omega_{\rm M}} M\frac{f_\ast }{t_\ast t_H(z)}\right]\frac{1}{M} \hs\frac{{\rm photons}}{M_{\odot}\hs {\rm s}}~ ,
\end{equation} 

\noindent where the quantity within the brackets is the SFR in a halo of mass $M$, $t_\ast$ is the mean star-formation time-scale in units of the Hubble time, $t_H(z)$, and $f_\ast$ is the fraction of baryons which is converted into stars.  For convenience, we combine the free parameters from eq. (\ref{eq:epsion}) into a single, redshift-independent, fiducial efficiency of ionizing photons per mass in stars, $\epsilon_{\rm fid}$, as:

\begin{align}
\label{eq:toteff}
\epsilon_{\rm ion}(z)\equiv \toteff\times 3.8 \times 10^{58}\left[\frac{\Omega_b}{\Omega_{\rm M}} \frac{1}{t_H(z)}\right]\hs\frac{{\rm photons}}{M_{\odot}\hs {\rm s}}.
\end{align}
\begin{align}
\toteff\equiv \left(\frac{2/3}{t_\ast}\right) \left(\frac{f_\ast}{0.2}\right) \left( \frac{f_{\rm esc}}{0.02} \right) \left( \frac{\epsilon_\gamma}{6.3\times10^{60}\Msun^{-1}}\right)
\end{align}

\noindent These model ingredients are highly uncertain at high redshifts. Hence we use this parametrization so that the reader can easily substitute his/her own preferred efficiency.  The above fiducial values were chosen since they provide a reasonably-good fit to high-redshift observations:

\begin{itemize}

\item The dependence above of the SFR $\equiv\frac{\Omega_b}{\Omega_{\rm M}} M\frac{f_\ast }{t_\ast t_H(z)}$, on the halo mass has been shown to simultaneously reproduce the observed luminosity functions of Ly$\alpha$ emitting galaxies (LAEs) \citep{Shimasaku06, Kashikawa06, DWH07, SLE07, McQuinn07LAE}, and the $z=6$ Lyman Break galaxies (LBGs) \citep{Bouwens06} fairly well, assuming similar semi-analytic models \citep[see Fig~7 of][]{DW07}.

\item  The fiducial value of $f_{\rm esc}=0.02$ corresponds to the mean of the probability distribution function of $f_{\rm esc}$ that was derived from the cumulative HI-column density distribution along sightlines to (long-duration) $\gamma$-ray burst (GRB) afterglows \citep{CPG07}. This mean value was also confirmed in the numerical simulations of \citet{GKC08}.

\item The number of ionizing photons per mass in stars is $\epsilon_\gamma = 6.3\times10^{60} \Msun^{-1}$ or equivalently $n_\gamma \equiv \epsilon_\gamma \times \mu m_p \approx 6400$ ionizing photons per baryon, if one assumes a Salpeter IMF and a gas metallicity of $Z=0.04Z_{\odot}$ \citep{Schaerer03}.  We point out that for this choice of IMF, the number of ionizing photons emitted per baryon that is involved in star formation depends on gas metallicity $Z$ as $n_\gamma \sim 10^{4.2-0.0029[\log_{10}Z+9]^{2.5}}$ \citep{Schaerer03}.
\end{itemize}

Finally, the units of the flux $\flux$ are 'ionizing photons s$^{-1}$ pcm$^{-2}$'. On the other hand, the ionizing background is commonly quantified in terms of its mean intensity $J(\nu)$ which has the units 'erg s$^{-1}$ pcm$^{-2}$ sr$^{-1}$ Hz$^{-1}$'. If the spectrum of the ionizing background radiation field is assumed to be of the form $J(\nu)=J_{\rm 21}(\nu/\nu_H)^{-\alpha}\times 10^{-21}$ erg s$^{-1}$ Hz$^{-1}$ pcm$^{-2}$ sr$^{-1}$, then the $J_{21}$ value in the cell located at $({\bf x}, z)$ is trivially obtained from $\flux$ through the equation  

\begin{equation}
\label{eq:fluxSED}
 J_{\rm 21}({\bf x}, z)= \flux \times \frac{h_{\rm p}\alpha/(4\pi)}{10^{-21}\hs{\rm erg}\hs{\rm s}^{-1} \hs{\rm pcm}^{-2}\hs{\rm Hz}^{-1}\hs{\rm sr}^{-1}} ~ .
\end{equation} 

\noindent 

To explicitly show the dependence of our calculated $ J_{\rm 21}({\bf x}, z)$ on the fiducial efficiency choice $\toteff$, some of our results below will be presented in terms of $J_{21}/\toteff$. This allows the reader to substitute his/her preferred value of $\toteff$ and obtain the corresponding physical $J_{\rm 21}({\bf x}, z)$.

In this study, all calculations assume that only halos with masses $M\gsim1.7 \times 10^8 \Msun$ contribute to the generation of the ionization and flux fields. This mass at $z=7$ roughly corresponds to the minimum temperature facilitating efficient atomic cooling, $\Tvir=10^4$ K. However, since this mapping is redshift-dependent, this same virial temperature corresponds to somewhat lower halo masses at higher redshifts: $M=1.4 \times 10^8$, $8.6 \times 10^7$, and $6.0 \times 10^7$ $\Msun$ at $z=$ 7, 10, and 13, respectively.  Because of this, and since this atomic cooling cut-off is approximate, we assume a couple of different minimum masses when studying the impact of feedback on reionization.  Note that our fiducial, parametrized UVB distributions can be (fairly accurately) adjusted to accommodate different contributing minimum masses by adjusting $\toteff$ accordingly (i.e. by the factor of the ratio of the fraction of collapsed matter contained in halos of mass $M \gsim M_{\rm min}$ and our fiducial choice of $M \gsim 1.7 \times 10^8 \Msun$).
We stress again that the main goal of this paper is not to precisely model UV flux distributions at high redshifts (such a thing being impossible given our current modest knowledge/inferences of high-$z$ sources); instead we explore a wide parameter space to attempt to approximately answer how likely is it for radiative feedback to delay the advanced stages of reionization when $\Tvir \gsim 10^4$ K sources dominate the ionizing photon budget. Uncertainties of order unity will thus not deter us from this noble task.

\begin{figure*}
\vspace{+0\baselineskip}
{
\includegraphics[width=0.24\textwidth]{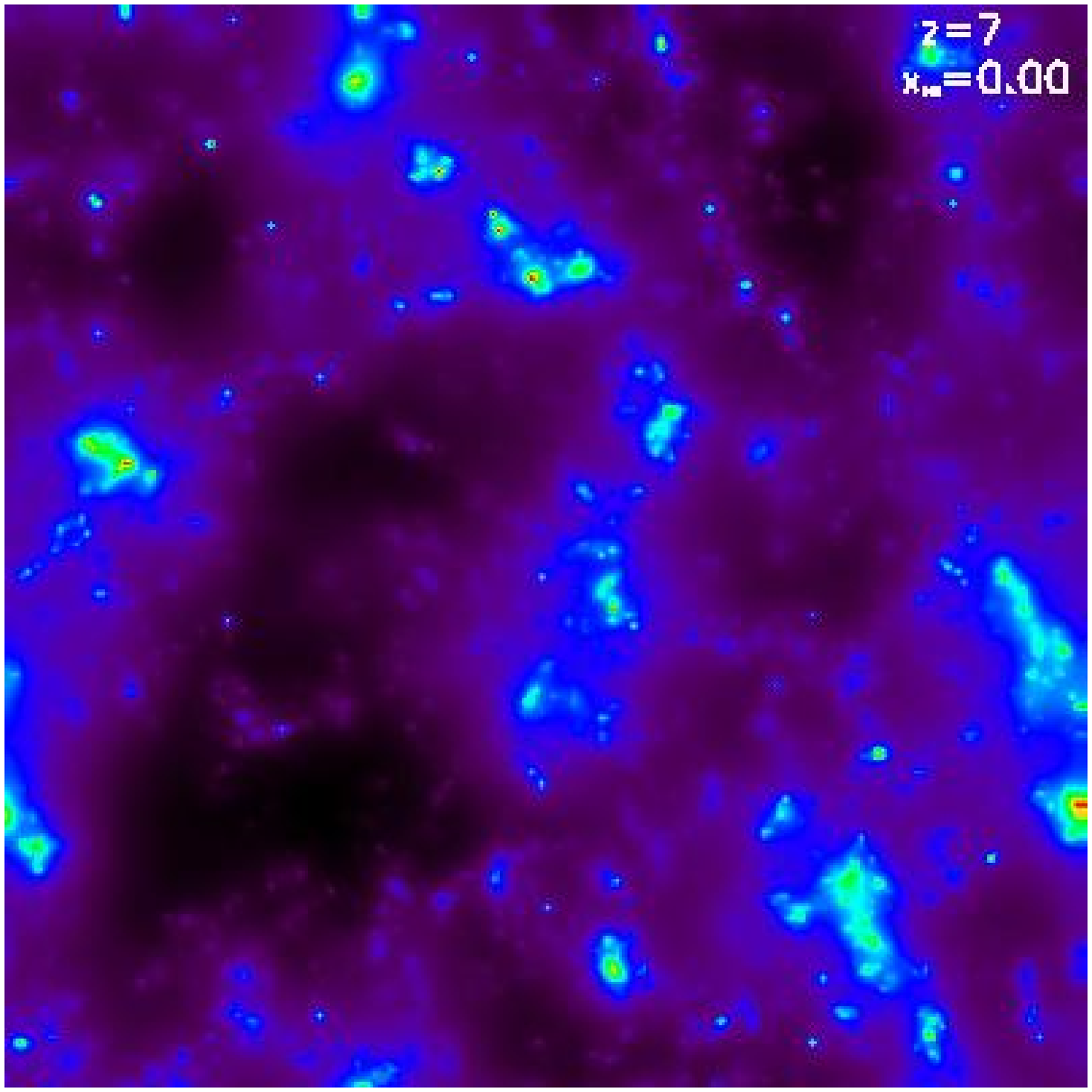}
\includegraphics[width=0.24\textwidth]{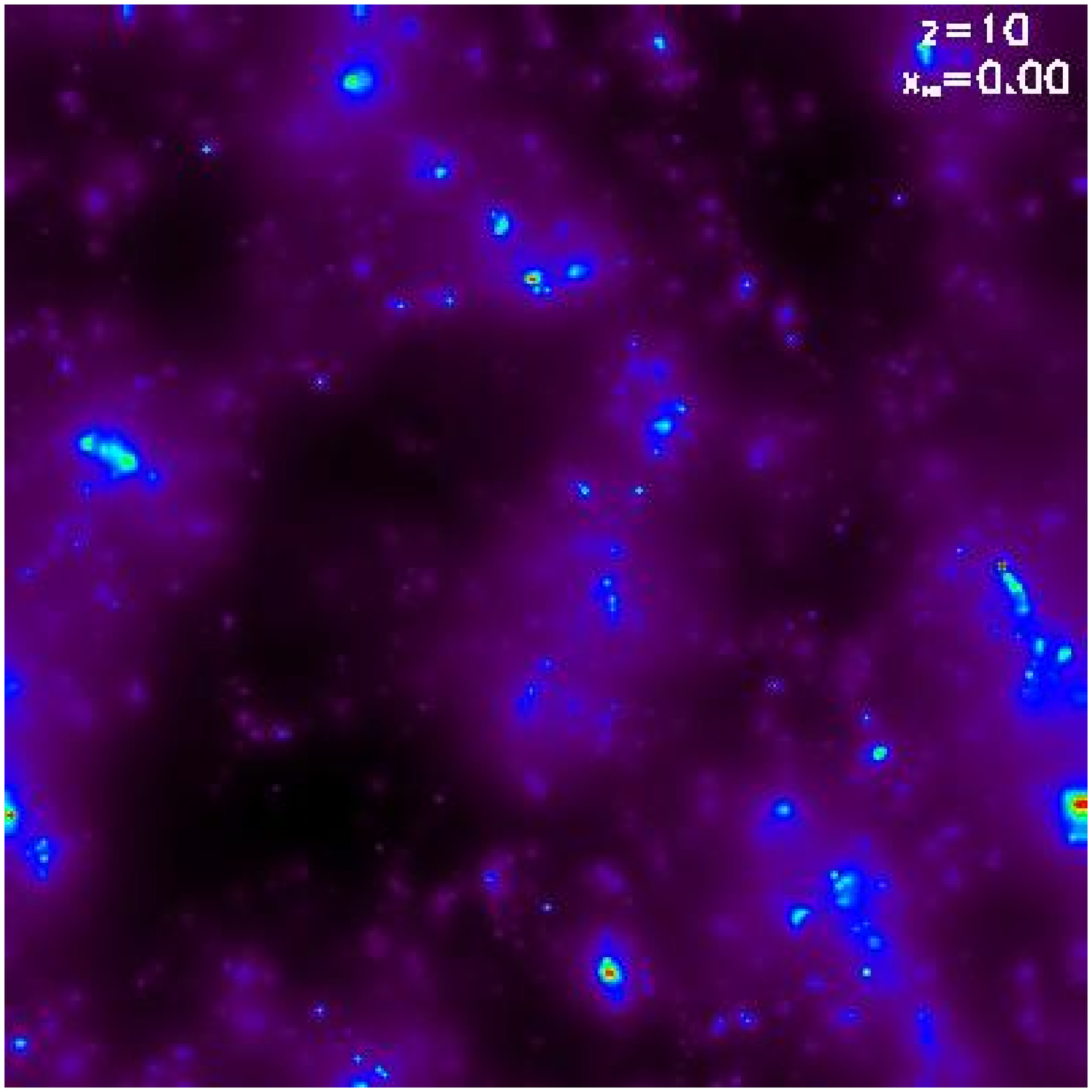}
\includegraphics[width=0.24\textwidth]{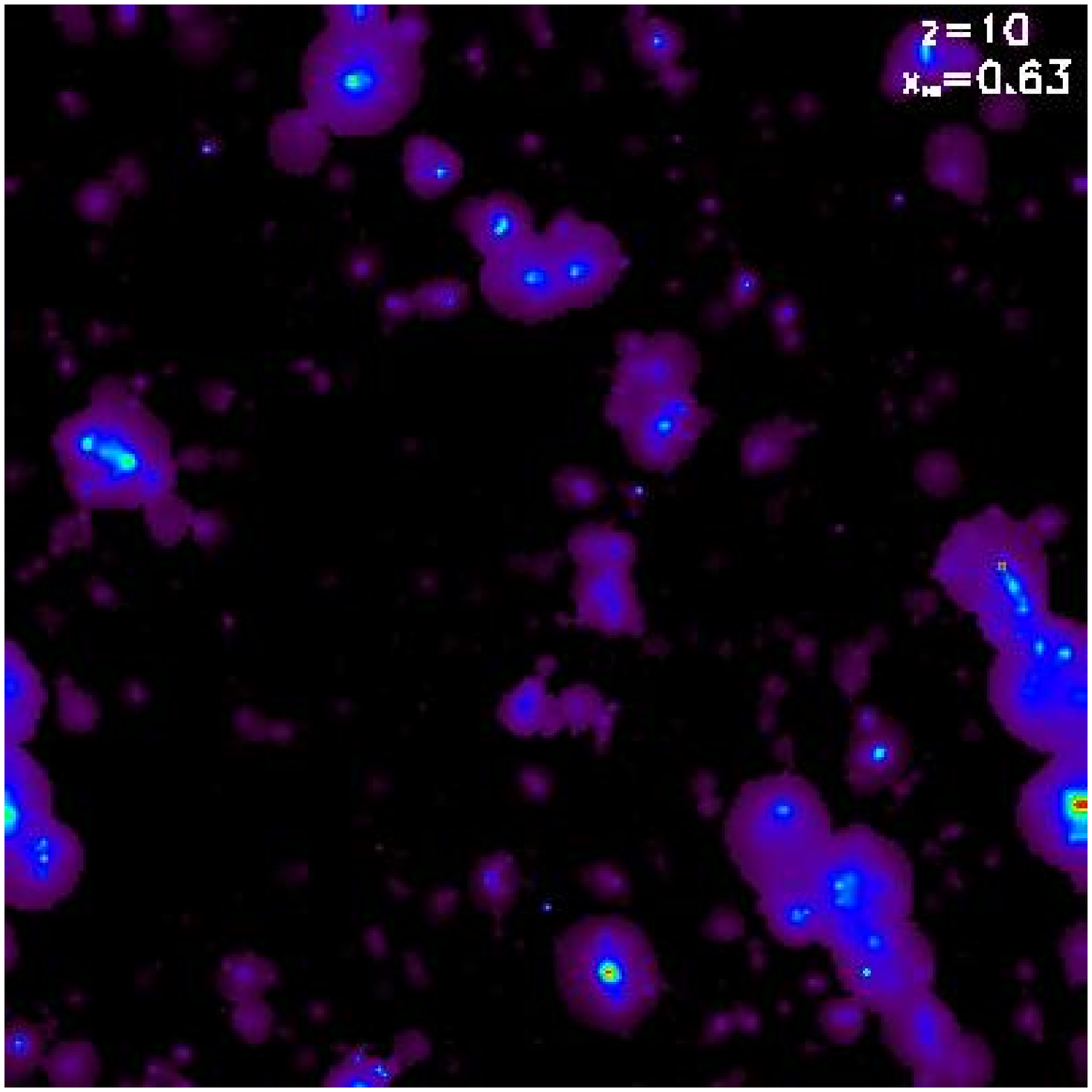}
\includegraphics[width=0.24\textwidth]{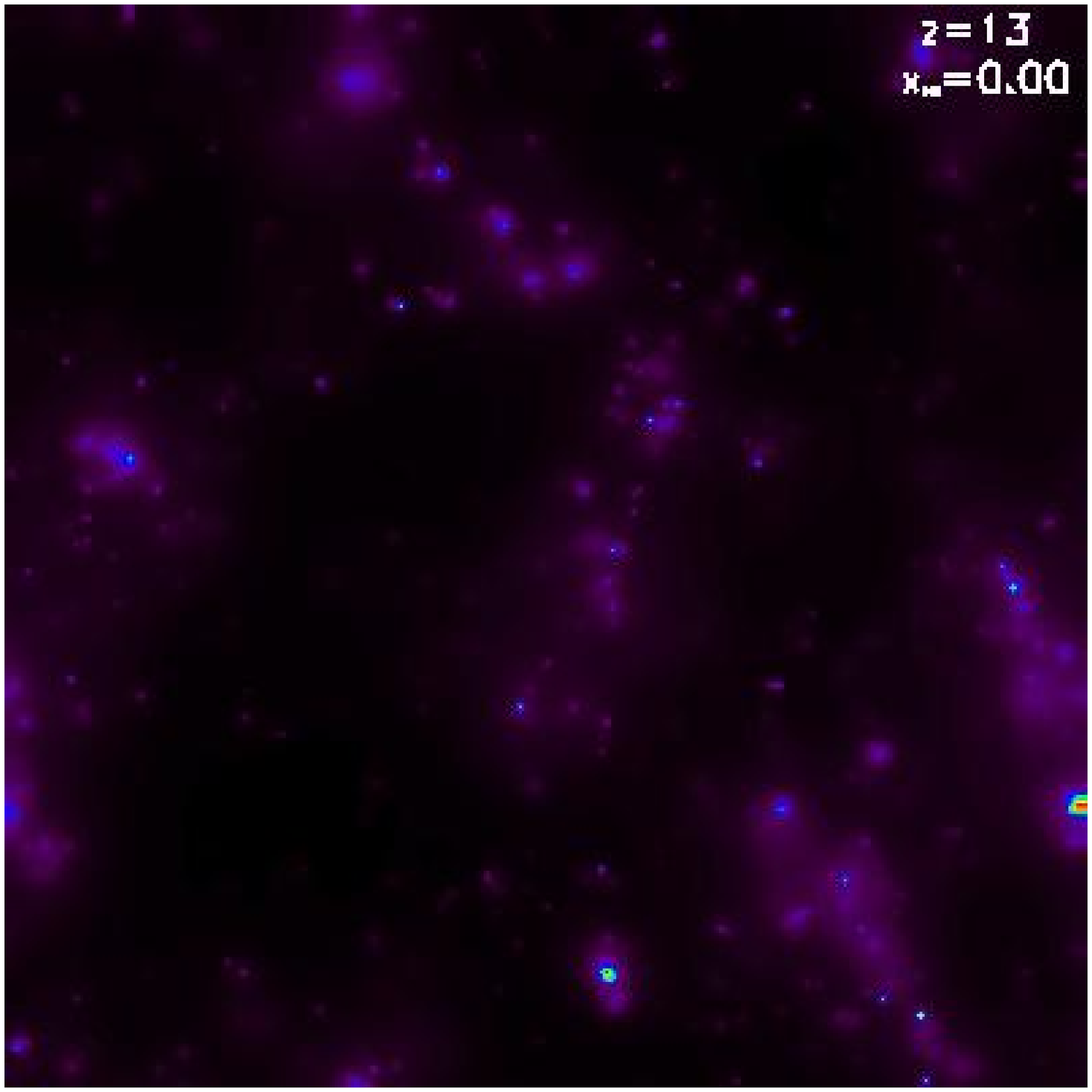}
\includegraphics[width=0.98\textwidth]{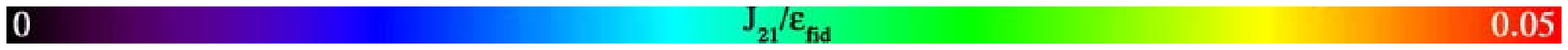}
}
\caption{
Slices through our UV flux boxes shown on a linear scale, scaled to the same maximum value.  Slices correspond to $(z, \avenf)$ = (7, 0), (10, 0), (10, 0.63), (13, 0).   Each slice is 0.5 Mpc thick and 100 Mpc on a side.
\label{fig:pics}
}
\vspace{+1.5\baselineskip}
\end{figure*}

In Figure \ref{fig:pics}, we present slices through several of our UV flux boxes, generated with the above procedure. The slices are shown on a linear scale, scaled to the same maximum value.  
The impact of the evolution of structure from $z=$ 13 to 7, and the 
truncation of the flux fields in neutral patches of gas, are qualitatively evident in these panels.

It is important to note that we treat the global neutral fraction $\avenf$, redshift $z$, and the source ionizing efficiency rate per mass $\epsilon_{\rm ion}$, all as separate free parameters.  We can afford this luxury because of the speed and versatility of our semi-numerical approach, as well as the confidence that our resulting ionization field topologies provide a good match to those generated by cosmological radiative transfer simulations \citep{MF07}.  In numerical simulations of reionization, the source efficiencies are coupled to the neutral fraction at redshift $z$.  Since we are not modeling the progress of reionization in this work (such a task being impossible at present), we keep $\epsilon_{\rm ion}$ and $\avenf$ separate, the former quantifying the ``{\it instantaneous}'' ionizing background at $z$, and the later depending on a complicated {\it history} of star formation and feedback.  Again, we do this in order to be as general as possible and explore a very wide swath of parameter space.

\section{Results}
\label{sec:results}

\subsection{Gas Collapse Fractions}
\label{sec:fcoll}

\begin{figure*}
\vspace{+0\baselineskip}
{
\includegraphics[width=0.33\textwidth]{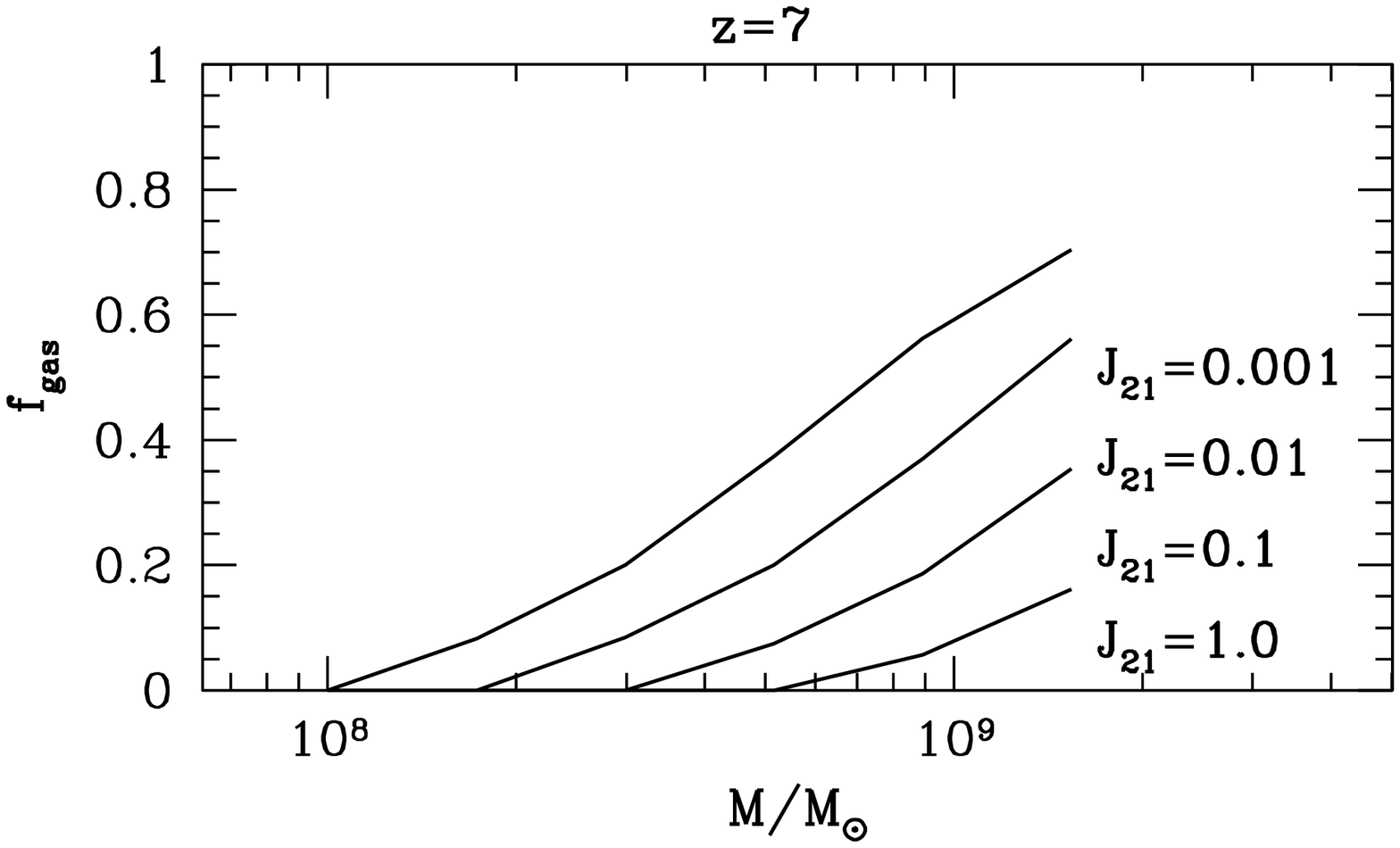}
\includegraphics[width=0.33\textwidth]{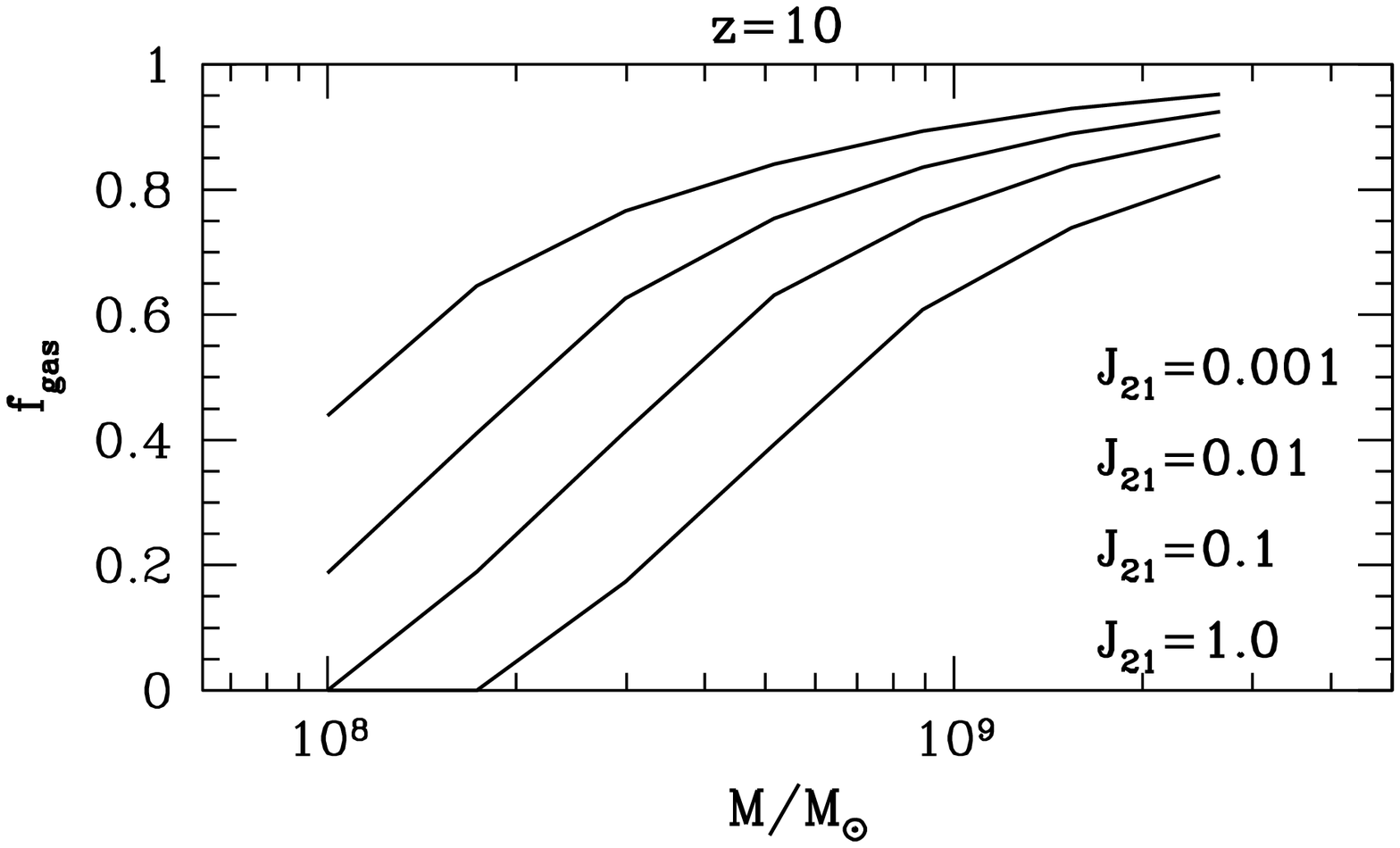}
\includegraphics[width=0.33\textwidth]{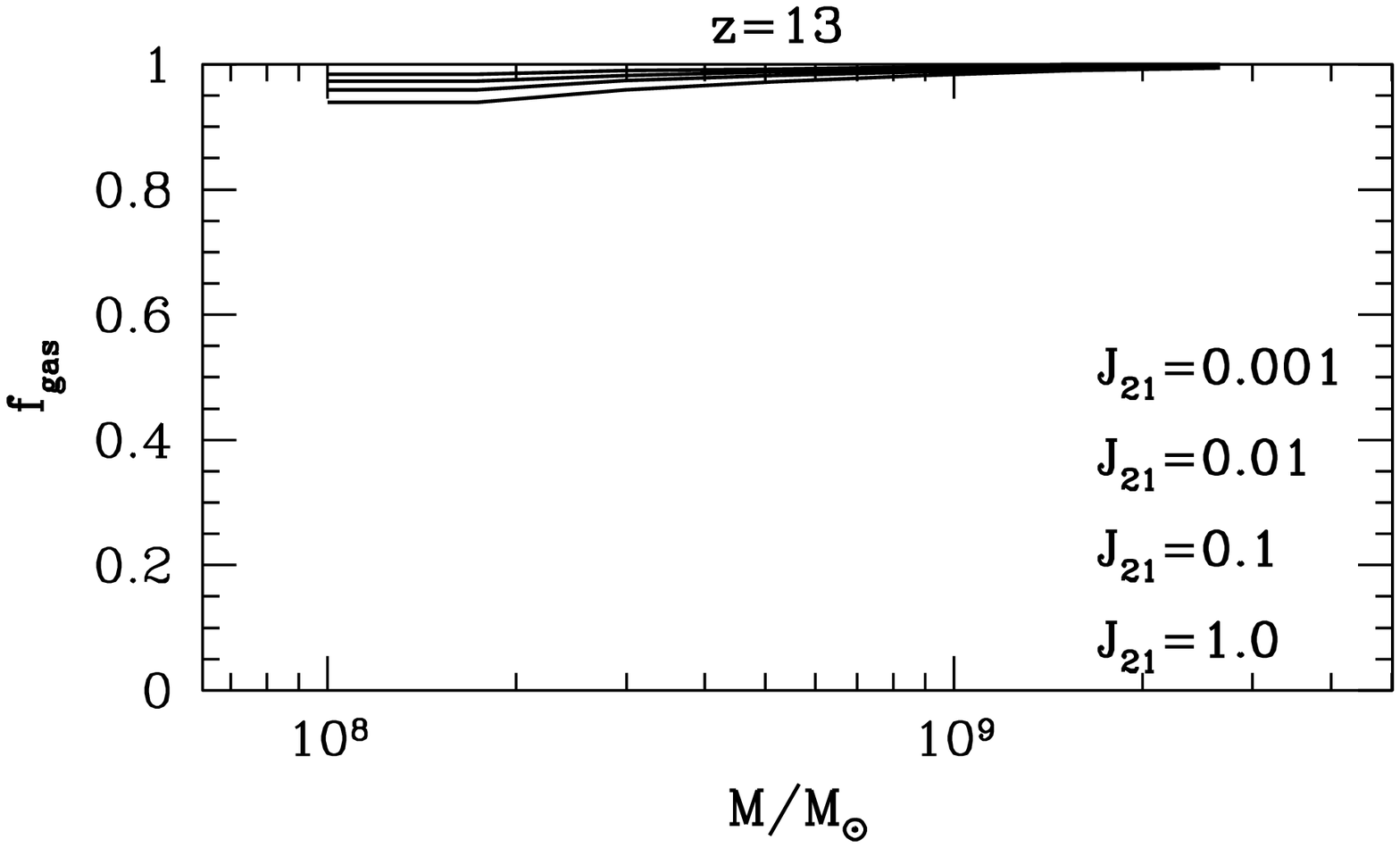}
}
\caption{
Fraction of gas (normalized to the $J_{21}=0$ case) which collapses into halos of mass $M$ at $z=$ 7, 10, 13 ({\it left to right panels}).  Curves correspond to $J_{21}=$ 0.001, 0.01, 0.1, 1.0, top to bottom. All runs assume $\zon=14$.
\label{fig:fcoll}
}
\vspace{+1.5\baselineskip}
\end{figure*}

In Figure \ref{fig:fcoll}, we present some results from the spherically symmetric collapse simulations.  The fraction of the gas (normalized to the case with no UVB)\footnote{The parameter $\fcoll \equiv M_{\rm gas}(p)/M_{\rm gas}(p=0)$, was defined as the mass of collapsed gas (i.e. virialized gas) divided by the mass of gas which would have collapsed in the absence of pressure.  The latter was found to closely resemble the $J_{21}=0$ case.}
 which collapses into halos of mass $M$ at $z=$ 7, 10, 13 ({\it left to right panels}) is shown as a function of the halo mass.  Curves correspond to $J_{21}=$ 0.001, 0.01, 0.1, 1.0.  The figures show that the amount of suppression is a strong function of the time elapsed since the halo started being exposed to the UVB \citep{Dijkstra04}, with even $10^8 \Msun$ halos at $z=13$ being too evolved at $\zon=14$ to lose more $\sim20\%$ of their gas to negative feedback even for fluxes as high as $J_{21}=100$ (not shown in the Fig.).

Since our primary interest in this work is how reionization is effected by radiative feedback, we will be especially interested in the critical specific flux values, $\Jcrit$, required to {\it entirely}\footnote{We define ``entirely'' based on the resolution of our collapse simulations, which roughly corresponds to the halo having $\lsim$0.2\% of the gas it would have had without being exposed to a UVB.  However our results are insensitive to this particular choice, as the collapsed gas fraction drops extremely rapidly with flux in the regime $\fcoll\lsim10$\%.} suppress the collapse of gas onto halos of mass $M$ at $z$.  We consider this to be more relevant to the issue of reionization than the fractional suppression of gas seen in Fig. \ref{fig:fcoll}. This is because star formation likely occurs on a longer time-scale than the time remaining for reionization to be completed once a large fraction of the universe is ionized and the $\Tvir \gsim10^4$ halos begin dominating the photon budget\footnote{Note that on average star formation should occur on some sizable fraction ($t_\ast =2/3$ in our model) of the hubble time, $t_H(z)$. At $z=10$, $(2/3) t_H = 500$ Myr; this time-scale is larger than, for example, the time elapsed from $z=10$ to $z=7$, $\Delta t = 300$ Myr.}
 (see also for example, Fig. 9 in \citealt{MJH06}, Fig. 2 in \citealt{HB06} and Fig. 1 in \citealt{Lidz07}).  The gas which was prevented from collapsing in Fig. \ref{fig:fcoll} was in the outer regions of the halo, and would likely have formed stars later, probably after reionization was completed.  Halos which retain even a little of their gas reservoir can form stars fairly quickly, as this gas is close to the high density core where cooling is efficient.  Thus the contribution of halos to the ``relative immediacy'' of completing reionization is better judged with $\Jcrit$.  However, since these assumptions are by no means certain, we present total feedback estimates in \S \ref{sec:feedback} based on two models: ({\it i}) partial gas suppression, in which we use the collapsed fraction, $f_{\rm gas}$, that is plotted in Fig. \ref{fig:fcoll}, and ({\it ii}) and total gas suppression, in which halos that see a UVB that is greater (less) than $J_{\rm crit}$ retain none (all) of their gas.

\begin{figure}
\vspace{+0\baselineskip}
\myputfigure{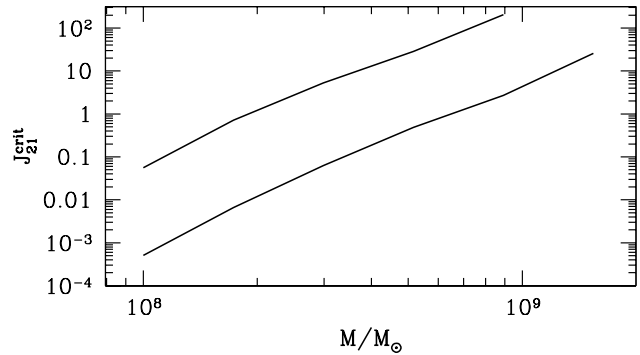}{3.3}{0.5}{.}{0.}
\caption{
$J_{21}$ values required to entirely suppress the collapse of gas onto halos of mass $M$ at $z=7$ ({\it bottom curve}) and $z=10$ ({\it top curve}).
\label{fig:MvsJcrit}}
\vspace{-1\baselineskip}
\end{figure}

In Figure \ref{fig:MvsJcrit} we show the critical $\Jcrit$ values required to entirely suppress the collapse of gas onto halos of mass $M$ at $z=7$ ({\it bottom curve}) and $z=10$ ({\it top curve}).  We obtain $\Jcrit$ at each $M$ by extrapolating the results from our collapse simulations linearly in $\fcoll$ vs. $\ln(J_{21})$. It is clear from the figure that $\Jcrit$ is a strong function of halo mass.  At this stage, one can already estimate that the contribution of $\sim10^8 \Msun$ halos to reionization could be suppressed in the later stages, but suppressing $\gsim$ few$\times 10^8 \Msun$ halos would be extremely difficult, even given our conservative assumptions.

\begin{figure*}
\vspace{+0\baselineskip}
{
\includegraphics[width=0.33\textwidth]{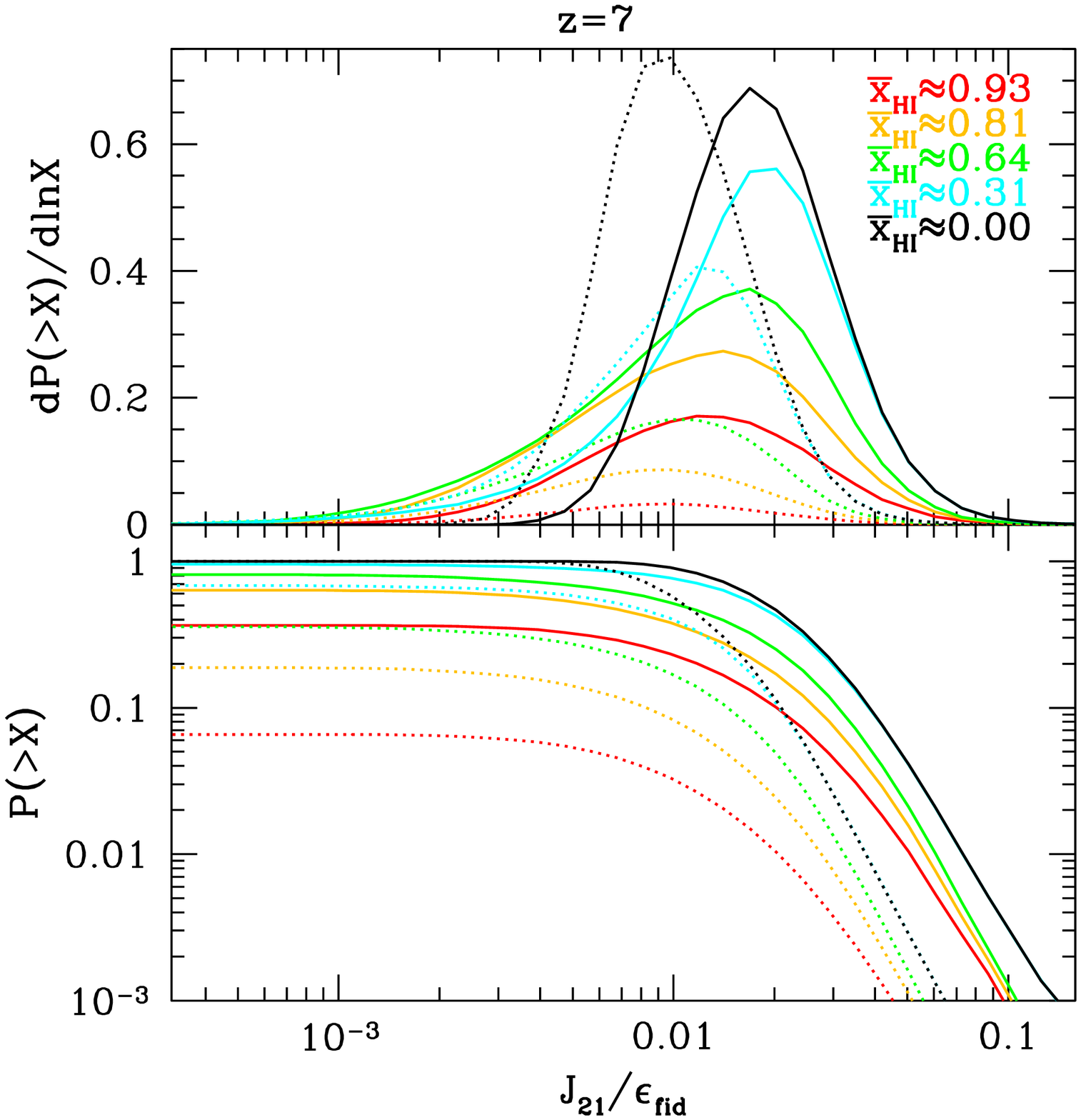}
\includegraphics[width=0.33\textwidth]{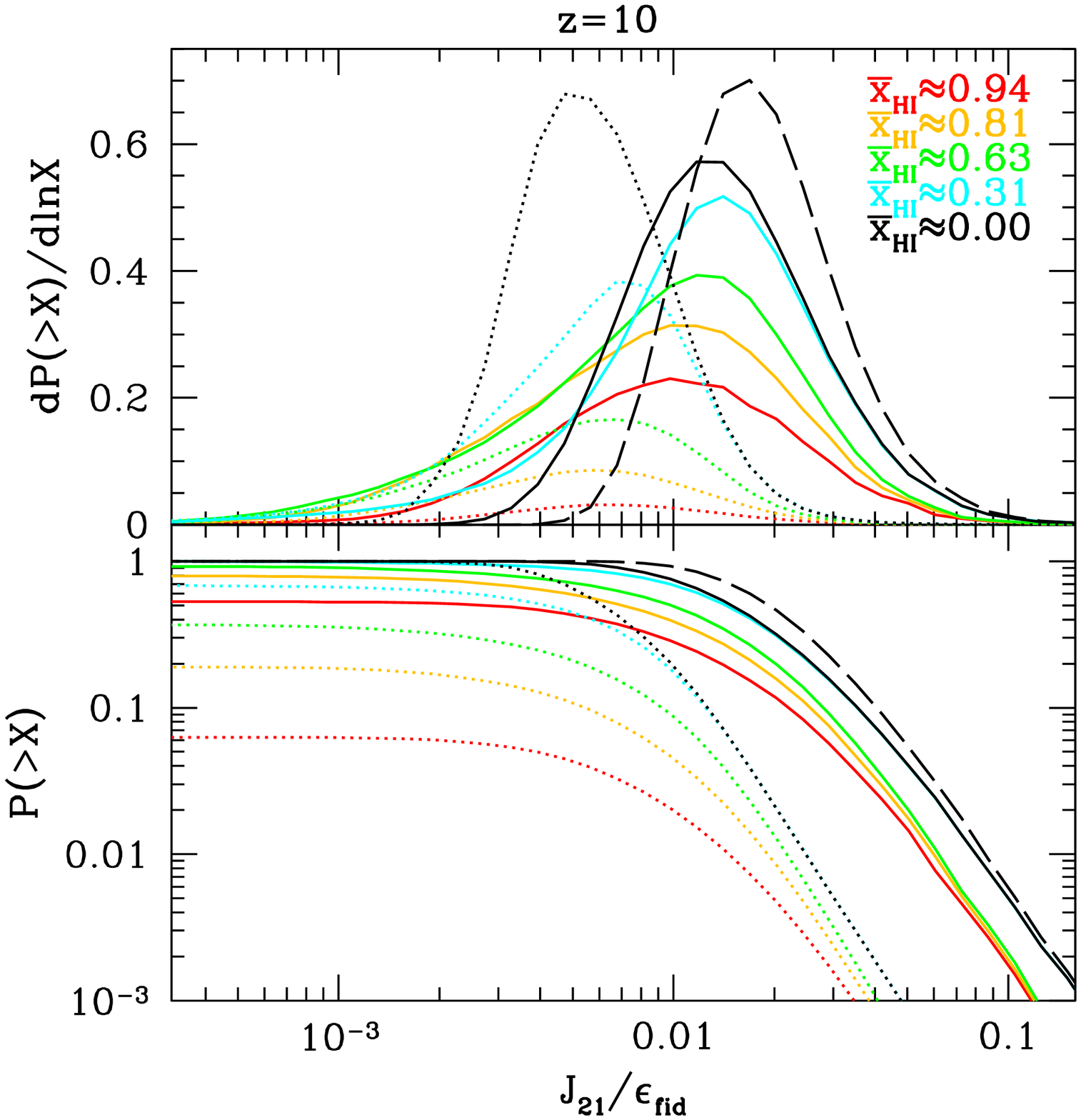}
\includegraphics[width=0.33\textwidth]{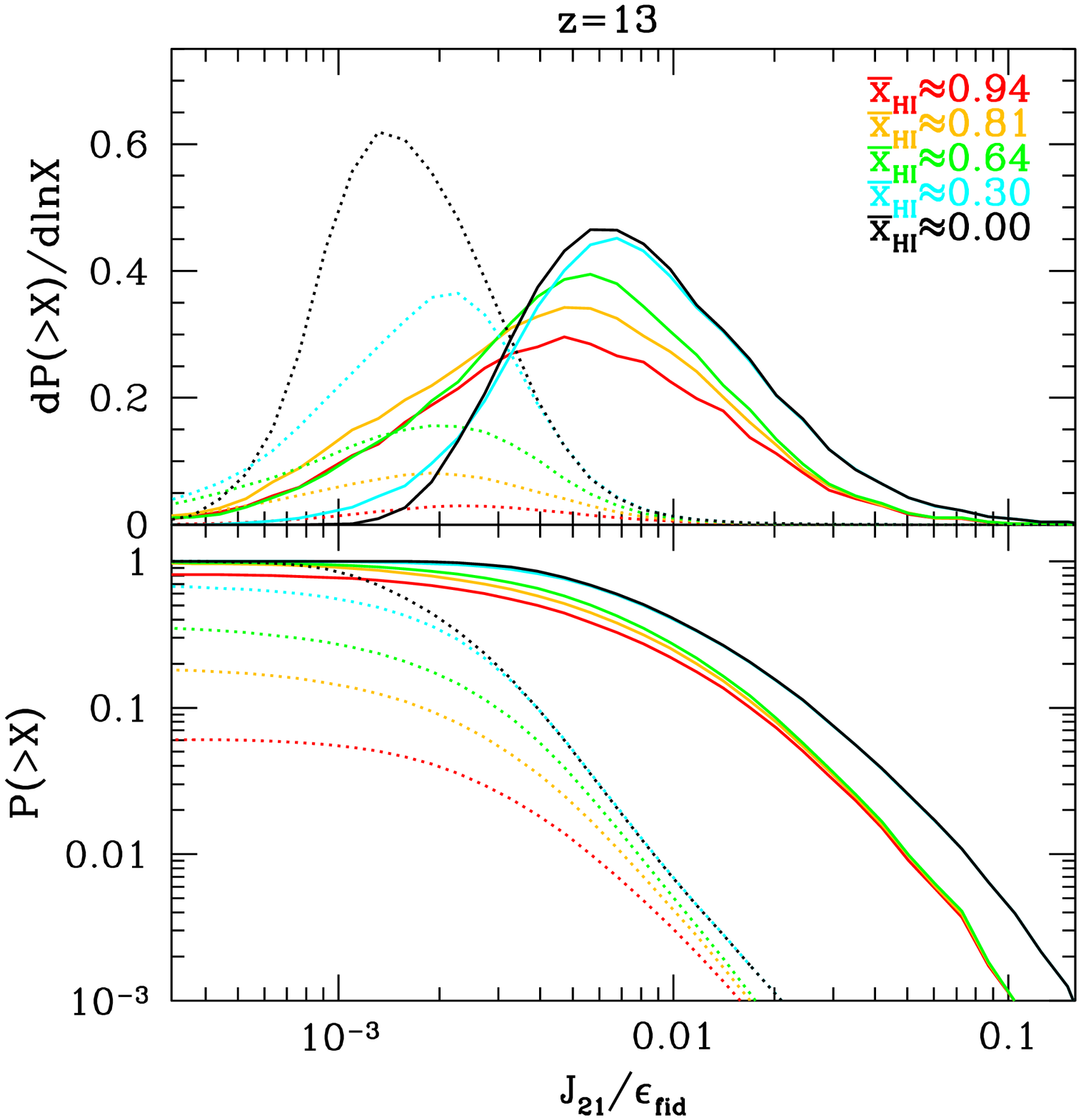}
}
\caption{
Probability density functions (PDFs; {\it top}) and cumulative probability density functions (CPDFs; {\it bottom}) of $J_{21}/\toteff$ at $z=$ 7, 10, 13 ({\it left to right}).  Dotted curves are generated from all pixels whereas solid curves are generated only from pixels which contain halos.  Curves correspond to $\avenf \approx$ 0, 0.31, 0.64, 0.81, 0.94 ({\it right to left}).  The dashed curves in the $z=10$ panel are generated from pixels which contain halos, assuming $\avenf\approx0$, but with a different m.f.p., $\lmfp=30$ Mpc.
\label{fig:CPDF}
}
\vspace{+1.5\baselineskip}
\end{figure*}

\subsection{UV Flux Distributions}
\label{sec:dist}

Following the procedure outlined in \S \ref{sec:flux}, we generate UV flux fields at $z=$ 7, 10, and 13, and for several values of $\avenf$.  Probability distributions of $J_{21}/\toteff$ extracted from these semi-numerical simulations are displayed in Figure \ref{fig:CPDF}. Dotted curves are generated from all pixels whereas solid curves are generated only from pixels which contain halos.  The dashed curves in the $z=10$ panel are generated from pixels which contain halos, assuming $\avenf\approx0$, but with a different m.f.p., $\lmfp=30$ Mpc.

Comparing corresponding dotted and solid curves clearly shows that halos are irradiated on average by an enhanced ionizing background, due to their clustering. This enhancement is larger at early times when sources are more biased.  Thus at $z=13$ the peaks of the distributions generated from halo locations are a factor of $\sim$4 larger than the peaks of the distributions generated from random locations in space, with this factor decreasing to $\lsim$1.5 by $z=7$.  One can also notice that as sources become less biased at lower redshifts, the flux distributions at low $\avenf$ become narrower and more gaussian. This is due to the fact that a larger fraction of the newly formed halos are born in less-biased regions with a correspondingly weaker UVB.  These new arrivals thus push the low-flux tail to higher values and narrow the distributions.

\begin{figure}
\vspace{+0\baselineskip}
\myputfigure{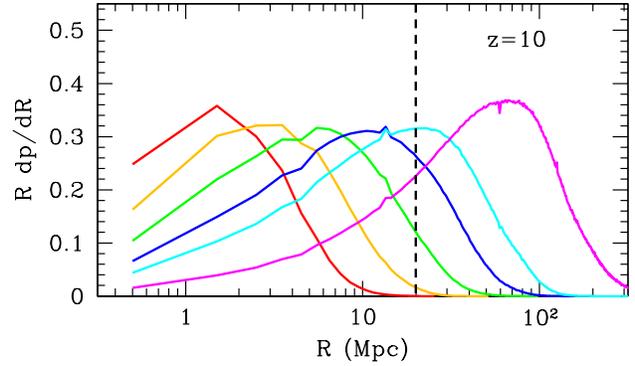}{3.3}{0.5}{.}{0.}
\caption{
Volume-weighted size distributions of ionized regions at $z=10$ and $\avenf \approx$ 0.94, 0.81, 0.63, 0.45, 0.31, 0.11 ({\it left to right}).  The vertical dashed line denotes our fiducial choice of $\lmfp=20$ Mpc.
\label{fig:size_dist}}
\vspace{-1\baselineskip}
\end{figure}

The imprint of our choice for the m.f.p., $\lmfp$, is also seen in the high-flux tails of the distributions in Fig. \ref{fig:CPDF}, especially in the solid curves.  There are two regimes governing the high-flux tail: (1) at low $\avenf \lsim 0.5$, the m.f.p sets the maximum flux seen by halos; (2) at high $\avenf \gsim 0.5$, the HII bubble size sets the maximum flux seen by halos.  Thus the high-flux tails of the distributions overlap at low $\avenf \lsim 0.5$ [regime (1)]; then at $\avenf \gsim 0.5$ the tails of the distributions drop and gradually decrease as the bubble size decreases and the halos see fewer and fewer sources [regime(2)]. The transition at $\avenf \sim 0.5$ denotes the moment when the characteristic bubble size (specifically the large tail of the distribution where the clustered sources lie) surpasses the $\lmfp$; this is confirmed in Fig. \ref{fig:size_dist}, where we plot the volume-weighted size distributions of ionized regions\footnote{These distributions were created by randomly choosing an ionized cell, and recording the distance from that cell to a neutral cell along a randomly chosen direction.  We repeat this Monte Carlo procedure 10$^7$ times for each $\avenf$. Note that the ionization topology and bubble distributions at fixed $\avenf$ are not sensitive to redshift over this interval \citep{McQuinn07, MF08LAE}.}.

Also shown in the middle panel of Fig. \ref{fig:CPDF} are the flux probability distributions ({\it dashed curves}) created from pixels which contain halos, assuming $\avenf\approx0$, but with a 50\% higher  m.f.p., $\lmfp=30$ Mpc.  By comparing these curves to the right-most solid curves, we can isolate the maximum impact of $\lmfp$, when bubble size is not an issue, i.e. $\avenf \approx 0$.  This increase in the m.f.p. shifts the peak of the flux distribution by $\sim$ 30\%; however, the shift is less pronounced at the high-flux than the low-flux end, since the ionizing radiation from nearby halos dominates over the large-scale background at the high-flux tail.  This, combined with the fact that the high-flux, high $\avenf$ tails overlap more at high $z$, when sources are more biased, suggests that the high-flux tails can effectively be modeled by a small-scale, Poissonian halo clustering component combined with a smooth contribution from larger scales (c.f. Dijkstra et al. 2008, in preparation).  Similarly, the high-flux tail of the UVB distributions might be extended if stocasticity in the source efficiencies is taken into account (Dijkstra et al. 2008, in preparation); however, in this paper, we study the impact on reionization, and any significant impact would have to result from fluxes not too far out on the high-value tail.

\begin{figure*}
\vspace{+0\baselineskip}
{
\includegraphics[width=0.45\textwidth]{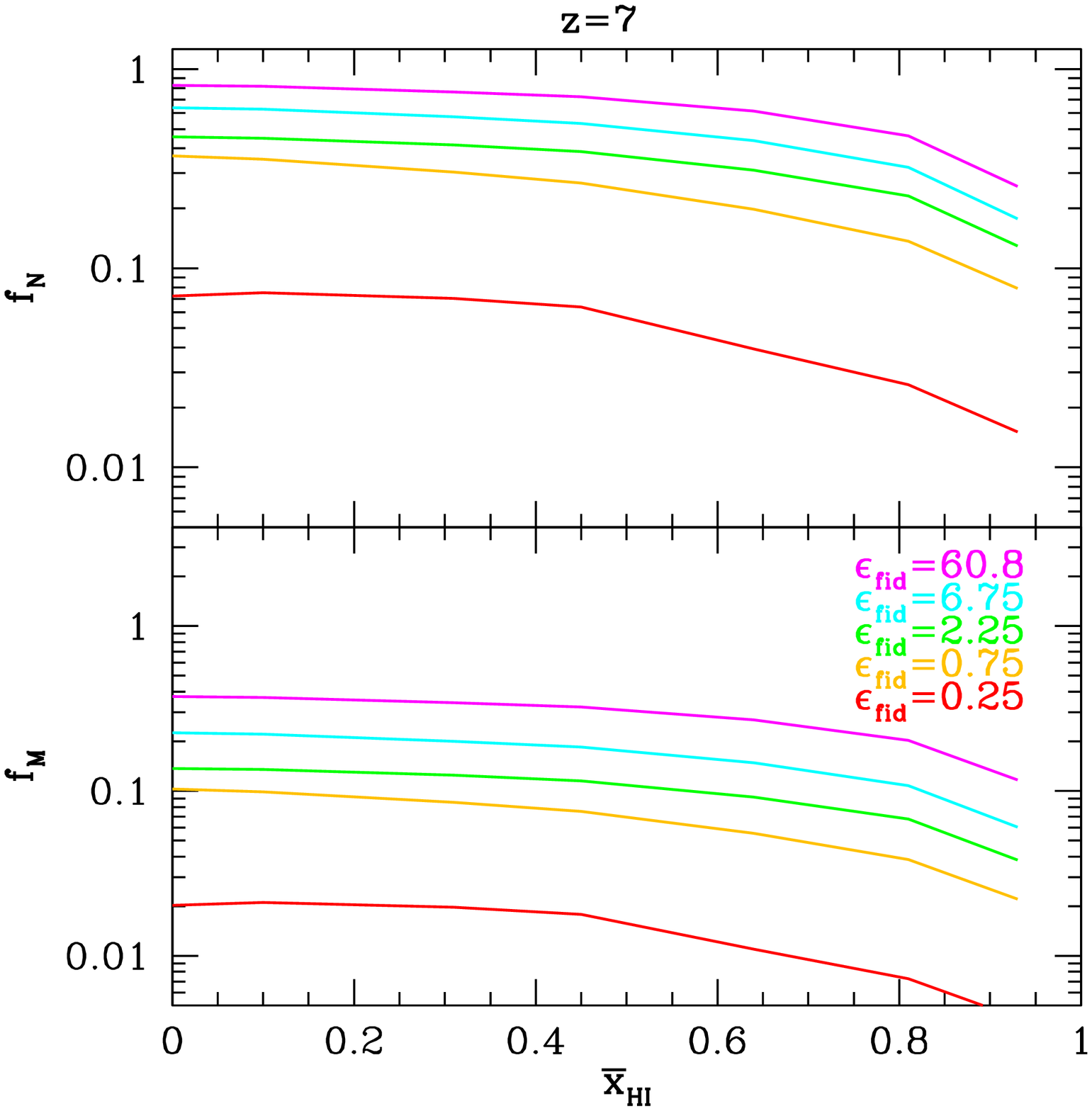}
\includegraphics[width=0.45\textwidth]{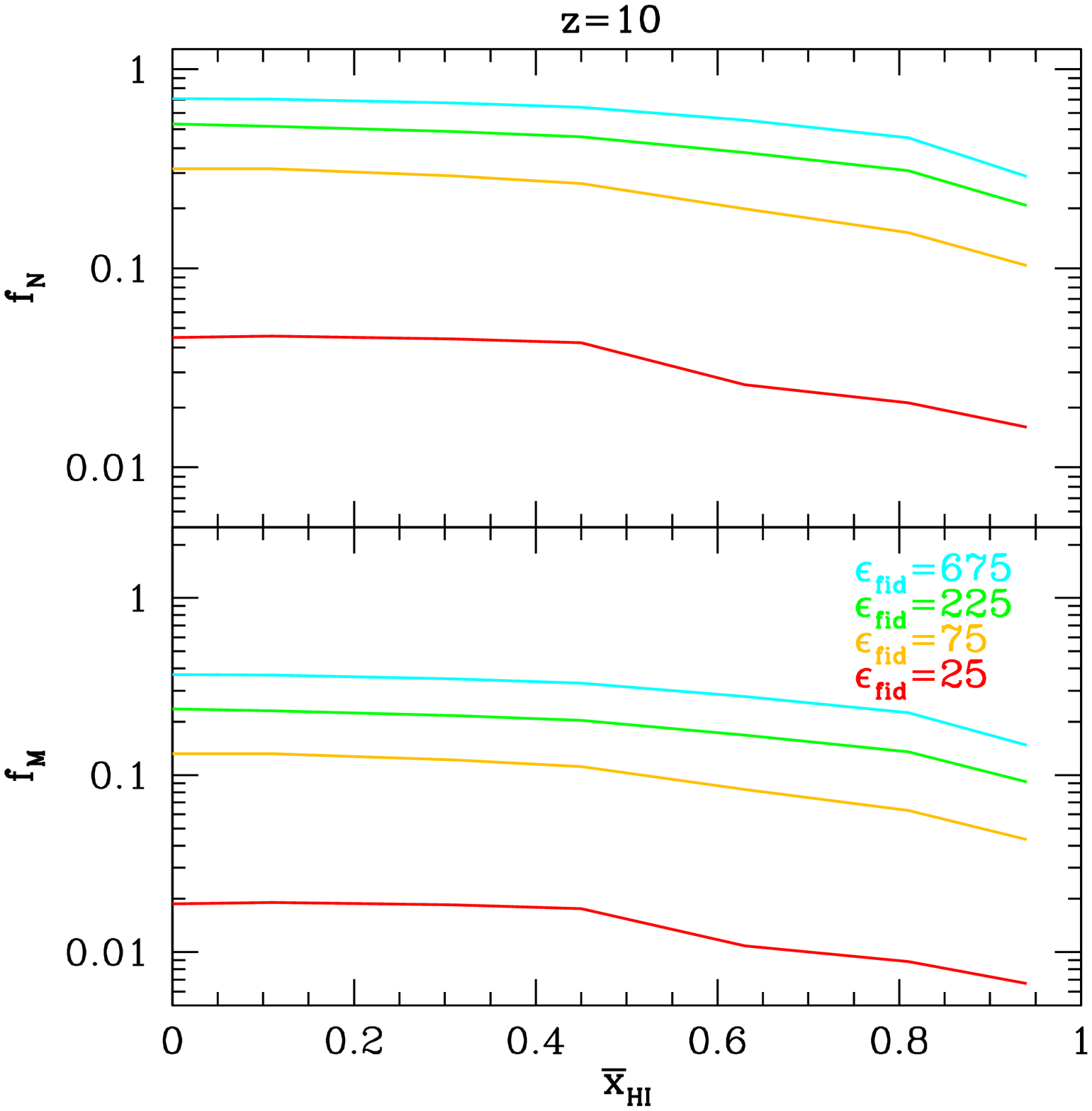}
}
\caption{
{\it Top: } Fraction of $M\ge1.7\times10^8 \Msun$ halos left without gas at $z=$ 7 and 10 ({\it left to right}). 
{\it Bottom:} Same as the top panels, but weighted by each halo's total mass. Curves correspond to $\toteff = $ 0.25, 0.75, 2.25, 6.75, 60.8 ({\it bottom to top in the left panels}) and $\toteff = $ 25, 75, 225, 675 ({\it bottom to top in the right panels}).
\label{fig:fsup}
}
\vspace{+1.5\baselineskip}
\end{figure*}

\subsection{UV Radiative Feedback}
\label{sec:feedback}

Combining our parametrized flux boxes with the results from our collapse simulation, we can proceed to answer the fundamental issue of this paper: can UV radiative feedback substantially 
suppress the contribution of $\Tvir\gsim10^4$ K halos to the advanced stages of reionization, when such halos are expected to dominate the ionizing photon budget?  We do this by counting the fraction of halos of mass $M$, which lie in regions of flux stronger than the critical flux, $\Jcrit$, shown in Fig. \ref{fig:MvsJcrit}, assuming various choices for our ionizing efficiency parameter, $\toteff$.  

In the top panels of Fig. \ref{fig:fsup}, we show the fraction of $M\ge1.7\times10^8 \Msun$ halos left without gas at $z=$ 7 and 10 ({\it left to right}).  The bottom panels show the same quantity, but weighted by each halo's mass.  These mass-weighted gas fractions are directly proportional to total gas mass 'locked up' in dark matter halos, which in turn determines the global star formation rate. The bottom panels therefore present a more realistic picture of the impact of UV feedback than the top panels.

From Fig. \ref{fig:fsup}, it is evident that ``reasonable'' choices of ionizing efficiencies, $\toteff\sim1$ are insufficient to quench star-formation in the majority of halos.  At $z=$ 7 (10) efficiencies of $\toteff\sim$ 5 (200) are required to completely evacuate/suppress gas from roughly half of $M \ge 1.7 \times 10^8 \Msun$ halos. The required efficiencies become even more out-of-reach when considering the mass-weighted gas fractions.  For example, suppressing more than half of the global gas reservoir requires $\toteff \sim$ 100 (1000) at $z=$ 7 (10)! In our scenario, halos at $z=7$ are much more susceptible to radiative feedback than those at $z=10$, since they have been exposed to an ionizing background radiation for twice as long (but note that we have exaggerated this effect, see \S~\ref{sec:mark_sims}).

We also note that the feedback effects are not particularly sensitive to the ionization efficiency, $\toteff$, once that parameter goes beyond a certain minimum value (e.g. $\toteff \gsim 1$ at $z=7$ or $\toteff \gsim 100$ at $z=10$).
This can be understood by noting that the critical flux curves in Fig. \ref{fig:MvsJcrit} are very strong functions of mass.  Once ionizing efficiencies are large enough that our smallest mass scale experiences feedback, additional large increases in $\toteff$ are required to significantly suppress gas collapse onto more and more massive halos.

One can note the footprint of $\lmfp$ in Fig. \ref{fig:fsup} as well. The rise in the curves becomes shallower at the lower neutral fractions, $\avenf\lsim0.5$, when the characteristic bubble size surpasses $\lmfp$, as discussed above.

\begin{figure}
\vspace{+0\baselineskip}
\myputfigure{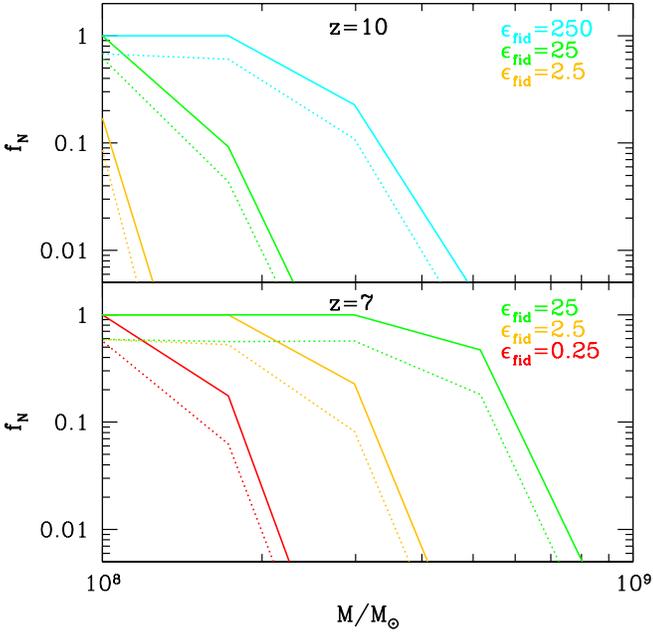}{3.3}{0.5}{.}{0.}
\caption{
Fraction of halos of mass $M$ left without gas at  $z=$ 7 and 10 ({\it bottom to top}).  Solid curves correspond to $\avenf\approx0$, while dotted curves correspond to $\avenf\approx0.81$. Each pair of curves assumes $\toteff = $ 250, 25, 2.5 ({\it right to left in the top panel}), and $\toteff = $ 25, 2.5, 0.25 ({\it right to left in the bottom panel}).
\label{fig:M_sup}}
\vspace{-1\baselineskip}
\end{figure}

It is useful to remove the mass function from the cumulative feedback effects depicted in Fig. \ref{fig:fsup}, and look at radiative feedback as a function of mass scale. Thus, in Fig. \ref{fig:M_sup} we plot the fraction of halos of mass $M$ left without gas at  $z=$ 7 and 10 ({\it bottom to top}).  Solid curves correspond to $\avenf\approx0$, while dotted curves correspond to $\avenf\approx0.81$.  We see that roughly an order of magnitude increase in the ionizing efficiency is required to extend the mass scale susceptible to strong feedback by a mere factor of $\sim2$.  Only $M\lsim2\times10^8 \Msun$ halos at $z=7$ experience significant feedback effects at our fiducial efficiency $\toteff\sim1$.

As discussed above, it is likely that estimates based on $\Jcrit$ are decent indicators of the strength of feedback effects during the advanced stages of reionization.  We justified this hypothesis in \S \ref{sec:fcoll} with a crude comparison of the probable time-scales of star formation and the completion of reionization.  This ``all or nothing'' scenario assumes that all halos of mass $M$ exposed to specific fluxes in excess of $\Jcrit$ no longer contribute ionizing photons, whereas those exposed to weaker fluxes continue doing so unhindered until the end of reionization.  
Nevertheless, it is useful to also consider a model with a less sharp cut-off due to feedback.  In this extreme case, we assume that the halos are able to ``rapidly'' (i.e. before the end of reionization) convert their entire gas reservoir at $z$ to ionizing radiation.  In this case, radiative feedback should follow the relative suppression of collapsed gas shown in Fig. \ref{fig:fcoll}.  Additionally, from a more fundamental standpoint, it is interesting to note the extent to which a UVB depletes the global gas reservoir at high-$z$.

\begin{figure}
\vspace{+0\baselineskip}
\myputfigure{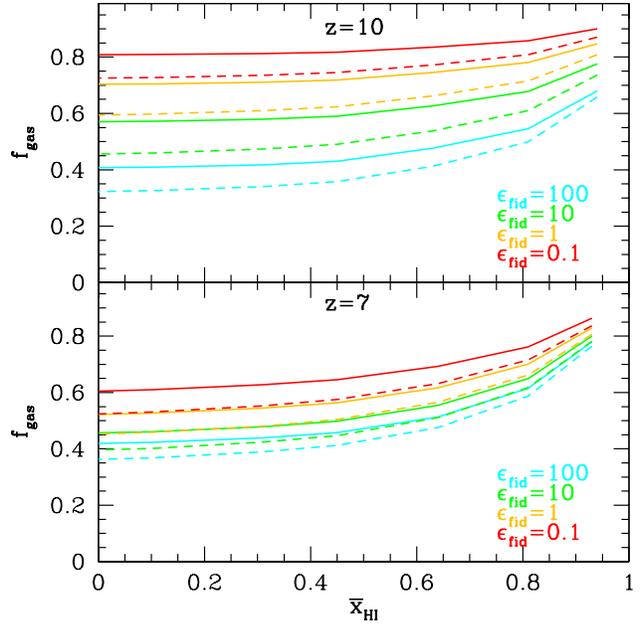}{3.3}{0.5}{.}{0.}
\caption{
Fraction of gas (normalized to the $J_{21}=0$ case) still remaining in $M>1.7\times10^8 \Msun$ ({\it solid curves}) and $M>10^8 \Msun$ ({\it dashed curves}) halos at $z=$ 7 and 10 ({\it bottom to top}). Pairs of curves correspond to $\toteff = $ 0.1, 1, 10, 100 ({\it top to bottom}).
\label{fig:gas_sup}}
\vspace{-1\baselineskip}
\end{figure}

Therefore, in Figure \ref{fig:gas_sup} we plot the fraction of gas (normalized to the $J_{21}=0$ case) still remaining in $M>1.7\times10^8 \Msun$ ({\it solid curves}) and $M>10^8 \Msun$ ({\it dashed curves}) halos at $z=$ 7 and 10 ({\it bottom to top}). Pairs of curves correspond to $\toteff = $ 0.1, 1, 10, 100 ({\it top to bottom}), at both redshifts.

  As in the top curves of Fig. \ref{fig:fsup}, feedback is not very sensitive to $\toteff$, requiring large increases in $\toteff$ to suppress gas in more and more massive halos.  This is especially evident at $z=7$, where feedback effects are strongest: here we see the remaining gas fraction only evolves $\lsim20$\% as $\toteff$ is increased from 0.1 to 100. An effective ``asymptote'' at $\fcoll\sim0.4$ is reached as the majority (though not all) of the gas is prevented from collapsing onto the ``vulnerable'', $10^8 \lsim M \lsim 10^9$ halos.  The suppressed fraction, $1-\fcoll\approx0.6$, roughly corresponds to the global collapse fraction of these halos at $z=7$.  This effective asymptote is at a lower $\fcoll$ at $z=10$, reflecting the smaller characteristic halo mass at that redshift and the corresponding higher global collapse fraction contained in these small, ``vulnerable'' halos.

\section{Assumptions and Uncertainties}
\label{sec:ass}

In \S~\ref{sec:mark_sims} we investigated the impact of a photoionizing flux on the ability of the gas to cool and collapse onto dark matter halos by performing 1-D hydrodynamical simulations. These calculations {\it overestimated} the impact of a UVB on the ability of gas to cool and collapse into the centers dark matter halos for a number of reasons:

\begin{itemize}

\item We assumed the spectrum of the UVB to be hard: $J(\nu)\propto J_{\rm 21}(\nu/\nu_H)^{-\alpha}$ and $\alpha=1.0$, which is typically associated with accreting black holes. However, stellar sources are likely to significantly contribute to the UVB at these redshifts.  These sources have much softer spectra, with $\alpha\sim 5$ \citep[e.g.][p.70]{BL01}. Under the assumption that the gas is transparent to ionizing photons, the photoionization [photoheating] rate scales as $\propto J_{\rm 21}/(\alpha+3.0)$ [$\propto J_{\rm 21}/(\alpha+3.0)(\alpha+2.0)$]. Choosing a hard spectrum is (almost) equivalent to boosting $J_{\rm 21}$, because in our simulations self-shielding is ignored and the gas {\it is} transparent to ionizing photons. Thus if the UVB is dominated by stellar sources at these high redshifts, we expect UVB intensities to be a factor of $\sim$ five lower than our conservative estimates (effectively $\toteff \sim 0.2$ for our reionization estimates)!

\item Our simulations ignore that at sufficiently large densities ($\delta \gsim 100$, e.g. Yang et al, 2006), the gas is capable of self-shielding against the ionizing background \citep[also see e.g.][]{SU04}. The reasons for ignoring self-shielding in our calculations are ({\it i}) it speeds up the simulations: properly accounting for self-shielding requires one to compute the increasing hardness of the spectrum as one moves deeper into the cloud, and ({\it ii}) in more realistic 3-D models of collapsing gas clouds, ionizing photons may penetrate deeper into the cloud via paths of lower HI column densities than one expects from a 1-D calculation. Hence, if one accounts for self-shielding using a 1-D simulation, one may underestimate the impact of the UVB, while throughout this paper we have conservatively chosen to overestimate the impact of the UVB. For example, Dijkstra et al (2004) found that $\sim 95\%$ of the gas was able to collapse into dark matter halos with $v_{\rm circ}=15$ km s$^{-1}$ (roughly corresponding to $10^8 \Msun$) that collapsed at $z=11$ for $z_{\rm on}=17$ and $J_{\rm 21}=10^{-2}$ when self-shielding was included. This is a significant rise compared to the $\sim$10\% fraction that was found for a model in which self-shielding was ignored.

\end{itemize}

A large uncertainty in our determinations of the UVB is the ionizing photon mean free path inside ionized regions, $\lmfp$. Our choice of $\lmfp=20$ Mpc is consistent with other theoretical estimates (e.g. \citealt{FO05, CFG08}), and might even be conservatively large (i.e. lowering $\lmfp$ only serves to reduce the calculated values of the UVB) at $z\gsim9$ given the extrapolation of LLS density from lower redshift observations: $\lmfp \approx 110 [5/(1+z)]^3$ Mpc \citep{Miralda-Escude03}.
However, given that that the nature of these absorbers is unknown, all current estimates are just educated guesses so similarity among them is a somewhat underwhelming validation.  Nevertheless, we demonstrated that $\lmfp$ only affects the UVB when the characteristic bubble size (weighted by the halo occupancy) surpasses $\lmfp$. Therefore, if $\lmfp$ is larger, then we have {\it underestimated} the value of the UVB, but {\it only during the very late stages of reionization, when the characteristic bubble size exceeds $\lmfp> 20$ Mpc}. Hence, we conclude that our main conclusion that UV feedback does not delay the bulk of reionization is completely unaffected by the uncertainty in $\lmfp$.  
%On the other hand, it is still interesting to consider that a rapid rise\footnote{Such a rise could be due to, for example, the evaporation of minihalos, a decrease in the covering factor of absorbers due to cosmic expansion, progress of ionized regions into underdense regions with a smaller density of absorbers, etc.  Note again that $\lmfp$ is defined as the m.f.p only {\it inside ionized regions}; thus the mere growth of HII bubbles does not affect $\lmfp$.} in $\lmfp$ beyond $\sim$ 20 Mpc, and the associated increase in negative feedback,  might prolong the end stages of reionization. For this to happen, $\lmfp$ would have to rise {\it quickly enough} compared to the time remaining for reionization to be completed (i.e. the growth of structure not experiencing feedback), {\it yet slowly enough} compared the the timescale necessary for UV feedback to have an impact (based on this work, this is roughly the hubble time at these redshifts). Without this fine-tuning, reionization will complete before the increase in feedback ``sets-in''.

Although high-redshift sources remain poorly understood, our fiducial choice of source ionization efficiency, $\toteff$, was motivated by its ability to reproduce the observed luminosity functions of $z\sim6$ LAEs \citep{Shimasaku06, Kashikawa06}, and LBGs \citep{Bouwens06} fairly well, assuming similar semi-analytic models \citep{DWH07, SLE07, McQuinn07LAE}.  It seems that significantly delaying the bulk of reionization through UV feedback is only likely if ionizing efficiencies of $z\gsim10$ sources are much higher (a couple of orders of magnitude; see right panel of Fig. \ref{fig:fsup}) than such $z\sim6$ data seem to imply.  An factor of $\sim4$ can be squeezed-out from eq. (\ref{eq:toteff}) if one assumes all star formation occurs on a dynamical time, and an additional factor of $\sim10$ can be obtained if one assumes ionizing emissivity, $\epsilon_\gamma$, expected of Population III stars \citep{Schaerer03}.  This seems unlikely, although such a strong redshift evolution in ionizing efficiency is implied if the \citet{Stark07} lensed sample of $z\sim9$ LAEs is genuine \citep{SLE07, MF08LAE}. Furthermore, $\epsilon_{\rm fid}$ depends linearly on $f_{\rm esc}$, which we assumed to be $f_{\rm esc}=0.02$ (see \S~\ref{sec:flux}). To our knowledge, no observations exist that hint at significantly larger average values of $f_{\rm esc}$ \citep[e.g.][]{Inoue06,Shapley06,Siana07}. Theoretically, escape fractions of order unity ($f_{\rm esc}\sim 1$) only appear associated with minihalos that contain a single massive star \citep{Whalen04}.

Again, we point out that we do not model reionization and feedback self-consistently.  Our conclusions on the strength of feedback effects are thus conservative, in that results from {\it WMAP} suggest most halos would have been exposed to a UVB for shorter periods of time than our choice of $\zon=14$ implies (see discussion in \S~\ref{sec:mark_sims}).  We are also conservative in neglecting the build-up of the UVB from $\zon=14$, assuming a constant distribution calculated at $z$ throughout the entire interval $\zon \rightarrow z$.  On the other hand, as we point out above, we assume a minimum mass of $\Mmin\ge1.7 \times 10^8 \Msun$ when generating ionization and flux fields, roughly corresponding to $\Tvir=10^4$ K at $z=7$.  But the same virial temperature corresponds to a somewhat lower $\Mmin$ at higher redshifts.  If star formation is indeed allowed down to exactly $\Tvir=10^4$ K, the contribution of neglected halos with $M<1.7\times10^8\Msun$ can be approximated with a $\sim40\%$ increase in the effective efficiency $\toteff$ at $z=10$ (assuming the ionizing photon contribution traces the collapsed fraction, as was done in this work).  This uncertainty is smaller the other uncertainties in $\toteff$.

Even assuming our source ionizing luminosity prescriptions are realistic, we do not expect to accurately model the tails of the UVB distributions, due to the following effects:
\begin{itemize}
\item Cosmic variance on the scale of our simulation box, 100 Mpc, causes us to miss statistically rare regions.  
\item Our halo fields are generated on a 1400$^3$ grid and our velocity fields are generated on a 700$^3$ grid.  Thus we cannot model the halo locations to better than 100 Mpc/700 $\approx$ 0.14 Mpc.
%\footnote{Note that, for example, the {\it r.m.s.} mass fluctuation in spheres with volume (0.14 Mpc)$^3$ at $z=10$ is $\sig = 0.6$ and at $z=7$ is $\sig = 0.8$.  Our semi-numerical simulation is accurate in the linear and quasi-linear regimes, $\sig \lsim 1$, (Mesinger et al. 2008, in preparation).  We thus expect our halo fields to be accurate on scales above the resolution limit.}
  This means that the very rare, very close source configurations are missed by our code.
\item We don't include a ``duty cycle'' in our source prescriptions, assuming that all sources are ``turned-on''.  While our prescription in eq. (\ref{eq:epsion}) accounts for this effect on average by sweeping it under the effective $f_\ast$ or $f_{\rm esc}$ parameters, the duty cycle stocasticity can extend the tails of the UVB distributions (Dijkstra et al. 2008, in preparation).  A similar effect is also produced if the absorbers are treated as distinct objects, instead of merely accounting for their mean flux decrement, $\exp(-r/\lmfp)$, as in eq. (\ref{eq:sum}).
\end{itemize}

Nevertheless, the tails of the UVB distribution are not important in answering the pertinent question of ``does UV radiative feedback significantly affect the progress of the bulk of reionization''.  Ionizing fluxes close to the means of the UVB distributions are more relevant in answering this question.  Given our generalized parametrizations, we are confident our models are sufficiently accurate for these purposes.   Thus we do not expect to loose sleep worrying about the effects mentioned in the previous paragraph.

\section{Conclusions}
\label{sec:conc}

In this paper, we quantify the importance of UV radiative feedback during the middle and late stages of reionization.  Specifically, we concern ourselves with halos capable of atomic cooling, $\Tvir\gsim10^4$ K. We first run suites of spherically-symmetric halo collapse simulations using various values of $z$, $M$, and $J_{21}$. We then generate parametrized UV flux distributions at $z=$ 7, 10, and 13, using semi-numerical, large-scale simulations of halo and ionization fields.  Combining these two results, we estimate how efficient is radiative feedback at hindering the progress of reionization, during its advanced stages (when $\Tvir \lsim 10^4$ K halos become subdominant contributors of ionizing photons).  This tiered approach allows us to explore an extremely wide range of parameter space, which is necessary to make any robust conclusions, given our poor knowledge of the properties of high-$z$ sources.

We find that under all reasonably conservative scenarios, UV feedback on atomically-cooled halos is not strong enough to notably delay the bulk of reionization.  For fiducial choices of source ionizing efficiencies (calibrated to match $z\sim6$ LAE and LBG luminosity functions) and turn-on redshift, fewer than $40$\% of $\Tvir \gsim 10^4$ K halos are left without gas at $(z, \avenf)$ $\approx$ (7, 0); this number drops to $\sim10\%$ when the distribution is mass-weighted (as is more appropriate for estimating the global star formation rate). Suppressing more than half of such halos requires a factor of $\sim$ 3--4 increase in fiducial ionizing efficiencies at $(z, \avenf)$ $\approx$ (7, 0) and over two orders of magnitude increase for the same fraction at $(z, \avenf)$ $\approx$ (10, 0)!

We find that feedback is very strongly dependent on halo mass.  For example, at $(z, \avenf)$ $\approx$ (7, 0) and $\toteff=0.25$, the fraction of halos left without gas decreases from 1 to 0.2 as the halo mass is increased only from $10^8$ to $1.7\times10^8$ $\Msun$.  The fraction of affected halos only decreases as $\avenf$ is increased.

  Accurate quantitative estimates will have to wait for a break-through in our understanding of the UV emission properties of high-redshift sources, and their dependence on halo mass.  Nevertheless, the inability of photoionization feedback to delay the middle and late stages of reionization is compelling, especially given that we do not include radiative transfer effects which would only decrease the relevance of UV feedback (see \S~\ref{sec:ass} for a more detailed discussion).

It seems that delaying the advanced stages of reionization through UV feedback on $M\gsim10^8\Msun$ halos is only likely if ionizing efficiencies of $z\gsim10$ sources are much higher ($\sim$ two orders of magnitude) than $z\sim6$ data seem to imply.  An evolution in $\toteff$ could be obtained by changing the fiducial values in eq. (\ref{eq:toteff}), possibly as a result of a top-heavy IMF or a star-burst dominated epoch.  However, an increase in the fiducial efficiency, $\toteff$, by $\sim$ two orders of magnitude seems unlikely.
 
Our results also suggest that the natural time-scale for the bulk of reionization is the growth of the collapsed fraction contained in $\Tvir\gsim 10^4$ K halos.  The natural timeframe for half of the universe to be ionized in such a prescription, using a toy model
%\footnote{Note that when modeling the HII filling factor one has to make assumptions concerning the ionization efficiency of halos as a function of their mass and formation history, as well as the modeling of $\lmfp$ and its evolution (e.g. due to photoevaporation of minihalos \citealt{SIR04}).}
 for the evolution of the HII filling factor \citep{FL05, MJH06, HB06, Lidz07} is $\Delta z_{\rm re} \sim$ 2 -- 3, with a small filling factor tail extending to higher redshifts. This tail is governed by more complicated feedback on $\Tvir \lsim 10^4$ K objects.  Such a scenario would be consistent with results implied by both the {\it WMAP} optical depth measurements \citep{Dunkley08, Komatsu08} and the SDSS quasars \citep{MH04, WL04_nf, Fan06, MH07}, especially given the inhomogeneous nature of reionization and the associated difficulties in interpreting present data \citep{LOF06, BH07, Maselli07, BRS07, MF08damp}.  This scenario is also in agreement with recent observational claims that faint galaxies at or below current detection thresholds dominate the ionizing photon budget at $z\gsim6$ \citep{YW04b, YW04a, Bouwens06}.

Finally, our results underscore the importance of extended dynamical ranges when modeling reionization.  Simulations must be capable of resolving halos with mass $\gsim 10^8 \Msun$, even when modeling the late stages of reionization, while at the same time being large enough to capture HII regions several tens of megaparsecs in size.

\vskip+0.5in

We are grateful to Steven Furlanetto and Zolt\'{a}n Haiman for helpful comments on the manuscript. MD is supported by Harvard University Funds.

\bibliographystyle{mn2e}
\bibliography{ms}

\end{document}